 \documentclass[preprint2]{aastex}

\newcommand{\2}{{~\sc ii}}
\newcommand{\3}{{~\sc iii}}
\newcommand{\4}{{~\sc iv}}

\newcommand{\mic}{{\,$\mu$m}}

\usepackage{appendix}

\shorttitle{Cornell Atlas of Spitzer/IRS Sources}
\shortauthors{Lebouteiller et al.}

\begin{document}


\title{CASSIS: The Cornell Atlas of Spitzer/IRS Sources}


\author{V.\ Lebouteiller\altaffilmark{1,2}, D.\ J.\ Barry\altaffilmark{1}, H.\ W.\ W. Spoon\altaffilmark{1}, J.\ Bernard-Salas\altaffilmark{1,3}, G.\ C.\ Sloan\altaffilmark{1}, J.\ R.\ Houck\altaffilmark{1}, and D. W. Weedman\altaffilmark{1}}
\altaffiltext{1}{ Department of Astronomy and Center for Radiophysics and Space Research, Cornell University, Space Sciences Building, Ithaca, NY 14853-6801, USA}
\altaffiltext{2}{Laboratoire AIM, CEA/DSM-CNRS-Universit\'e Paris Diderot DAPNIA/Service d'Astrophysique B\^at. 709, CEA-Saclay F-91191 Gif-sur-Yvette C\'edex, France}
\altaffiltext{3}{IAS, B{\^a}t. 121, Universit\'e Paris-Sud, 91435 Orsay, France}
\email{vianney@isc.astro.cornell.edu}







\begin{abstract}
We present the spectral atlas of sources observed in low resolution with the \textit{Infrared Spectrograph} on board the $Spitzer$ Space Telescope.  More than 11\,000 distinct sources were extracted using a dedicated algorithm based on the SMART software with an optimal extraction (AdOpt package). These correspond to all 13\,000 low resolution observations of fixed objects (both single source and cluster observations). 
The pipeline includes image cleaning, individual exposure combination, and background subtraction. A particular attention is given to bad pixel and outlier rejection at the image and spectra levels. 
Most sources are spatially unresolved so that optimal extraction reaches the highest possible signal-to-noise ratio. For all sources, an alternative extraction is also provided that accounts for all of the source flux within the aperture.
CASSIS provides publishable quality spectra through an online database together with several important diagnostics, such as the source spatial extent and a quantitative measure of detection level. Ancillary data such as available spectroscopic redshifts are also provided.
The database interface will eventually provide various ways to interact with the spectra, such as on-the-fly measurements of spectral features or comparisons among spectra.
\end{abstract}


\keywords{Atlases, Catalogs, Infrared: general, Methods: data analysis, Techniques: spectroscopic}



\section{Introduction}

The Infrared Spectrograph (IRS; Houck et al.\ 2004)\footnote{The IRS was a collaborative venture between Cornell
University and Ball Aerospace Corporation funded by NASA through the
Jet Propulsion Laboratory and the Ames Research Center.} is one of three instruments on board the $Spitzer$ Space Telescope (Werner et al.\ 2004) along with the two photometers Infrared Array Camera (IRAC; Fazio et al.\ 2004) and Multiband Imaging Photometer for $Spitzer$ (MIPS; Rieke et al.\ 2004). The IRS performed more than 21\,000 observations over the cryogenic mission lifetime (2003 November 30 - 2009 May 15), corresponding to about 14\,000 distinct targets. The IRS observed between $\approx5.2$ and $\approx38.0$\mic\ in two low-resolution modules with a resolving power of $R\sim60-120$ ($\approx75$\%\ of the observations), and two high-resolution modules with $R\sim600$ (Table\,\ref{tab:modules}).

\begin{table*}[h]
\begin{center}
  \caption{Modules of the Spitzer/IRS. CASSIS currently includes only the data from the SL and LL modules.}
  \label{tab:modules}
  \begin{tabular}{c c c c c}
  \hline
  Module & Order(s) & $\lambda$ (\mic) & Aperture size ($\arcsec$) & Pixel size ($\arcsec$) \\
  \hline
  SL & 1 & 7.4 - 14.5 & 3.7$\times$57 & 1.8 \\
  SL & 2 & 5.2 - 7.7  & 3.6$\times$57 & 1.8 \\
  SL & 3 & 7.3 - 8.7 &  3.6$\times$57 & 1.8 \\
  \hline
  LL & 1 & 19.5 - 38.0 & 10.7$\times$168 & 5.1 \\
  LL & 2  & 14.0 - 21.3 & 10.5$\times$168 & 5.1 \\
  LL & 3  & 19.4 - 21.7 & 10.5$\times$168 & 5.1 \\
  \hline
  SH & 11-20 & 9.9 - 19.6 & 4.7$\times$11.3 & 2.3 \\
  \hline
  LH & 11-20 & 18.7 - 37.2 & 11.1$\times$22.3 & 4.5 \\
  \hline
  \end{tabular}
\end{center}
\end{table*}

Two  products are essential to ensure the legacy of the Spitzer/IRS data. Most importantly, the community should have  available the highest quality spectra for the simplest cases, i.e., isolated point-like or partially extended sources with negligible (or uniform) background emission. In addition, specific tools are required to analyze the most complex cases, i.e, extremely faint sources (few tenths of mJy), blended sources, and/or sources with non-uniform background emission. 

While existing data reduction and analysis software such as SMART (Spectroscopic Modeling Analysis and Reduction Tool; Higdon et al.\ 2004) and SPICE (Spitzer IRS Custom Extraction\footnote{http://irsa.ipac.caltech.edu/data/SPITZER/docs/dataanalysistools/tools/spice/}) already provide several extraction algorithms, the AdOpt algorithm which we developed within SMART significantly improves spectral extractions with the IRS by using a super-sampled point spread function (PSF; Lebouteiller et al.\ 2010). AdOpt provides the best possible signal-to-noise ratio (S/N) while being a flexible algorithm able to handle blended sources and complex background emission. One of the first  applications was the separation of a supernova and a galaxy nucleus located in the same slit (Fox et al\ (2010). 

 The Cornell Atlas of Spitzer/IRS Sources (CASSIS)  utilizes an automatic spectral extraction tool based on AdOpt to provide the complete sample of staring observations (not including mappings) with the two low-resolution modules \textit{Short-Low} and \textit{Long-Low} (hereafter SL and LL). CASSIS complements the existing ``post-BCD'' database\footnote{Post-BCD products are available at http://irsa.ipac.caltech.edu/applications/Spitzer/Spitzer/} from the Spitzer Science Center (SSC), by providing optimal extractions as well as multiple diagnostic tools  including source extent and extractions with alternative backgrounds. Special attention is given to the
identification/removal of artifacts, and to accurate background
subtraction. The latter is especially important for faint sources, 
whose flux densities represent only a few percent of the total
emission (mostly originating from zodiacal dust at IRS wavelengths). 
In the future the CASSIS atlas will include serendipitous sources as well as high-resolution spectra taken with the \textit{Short-High} and \textit{Long-High} modules. 
 
A dedicated website has been developed\footnote{http://cassis.astro.cornell.edu/atlas} to allow users to query for observations and search for spectra described by various parameters. The CASSIS atlas currently contains about 13\,000 low-resolution spectra corresponding to $\sim$11\,000 distinct sources. Table \ref{tab:stats} shows the number of observations in the CASSIS atlas for each scientific category. The first application of the CASSIS atlas is the study of 301 galaxies observed with the IRS and with the Infrared Astronomical Satellite (IRAS) (Sarsgyan et al. 2011). The CASSIS spectra were used to measure the 7.7\mic\ PAH flux density and the warm dust continuum to derive the infrared luminosity. 

In the present paper, we review the most important steps of the extraction pipeline which leads to CASSIS (Sections 2 to 6). Complete and updated documentation is accessible online\footnote{http://isc.astro.cornell.edu/Smart/CassisPipeline}. Products, diagnostics, and the online interface are described in Sections 7, 8, and 9.  Finally, we present a few examples and illustrations of the possibilities enabled by the CASSIS atlas (Section 10).

\section{General description of the data processing}\label{sec:general}

About 85\%\ of the IRS observations were performed in the ``staring'' mode which observes the spectrum of a given source at two  different nod positions along the slit\footnote{More information on the observing
modes can be found in the Instrument Handbook at  http://irsa.ipac.caltech.edu/data/SPITZER/docs/ }. 
The two nod positions are sufficiently 
far apart that subtracting one nod image from the other removes 
the background effectively, except in cases when a serendipitous 
source is located at the other nod position, or when there is 
complex background emission. Note that backgrounds can also be subtracted using detector images when the source is in the slit for the alternative order (e.g., nods in the LL1 slit when the source is in the LL2 slit produce LL1 images of background only). More details on the available background subtraction methods are given in Sect.\,\ref{sec:lowlevel}. In staring mode, the observation sequence is performed as follows:\\
\textit{order 2 nod 1 - order 2 nod 2 - order 1 nod 1 - order 1 nod 2}

For a given observation (called AORkey, or AOR; see the list of acronyms in the Appendix), the staring mode observes either a single position or multiple positions (the latter case being also referred to as ``cluster observation'' which is designated by a single AORkey). The remaining $\lesssim15$\%\ observations were performed as spectral mappings in which the slit is progressively stepped across a extended source. We do not as yet include mappings in CASSIS. 

The official SSC pipeline consists of two parts, the Basic Calibrated Data (BCD) pipeline and the post-BCD pipeline. The BCD pipeline reduces the raw detector images from the individual exposures and removes the electronic and optical artifacts, including dark current, droop effect, non-linearity, radhit (detection and flagging of cosmic ray events), jail-bar pattern, and stray light. The BCD pipeline also includes a flat-field correction. The steps performed by the BCD pipeline are relatively well understood, and we consider the BCD products as reliable inputs for spectral extraction. We refer to the IRS Instrument Handbook for more details on the BCD products. 

The post-BCD processing, which  transforms calibrated 2D detector images to 1D spectra, often requires a manual intervention or an algorithm specifically designed for a given observation. The steps involved in the post-BCD processing notably includes image cleaning, background subtraction, and spectral extraction.

\begin{figure*}
\centering
\includegraphics[angle=90,scale=0.55,clip=true]{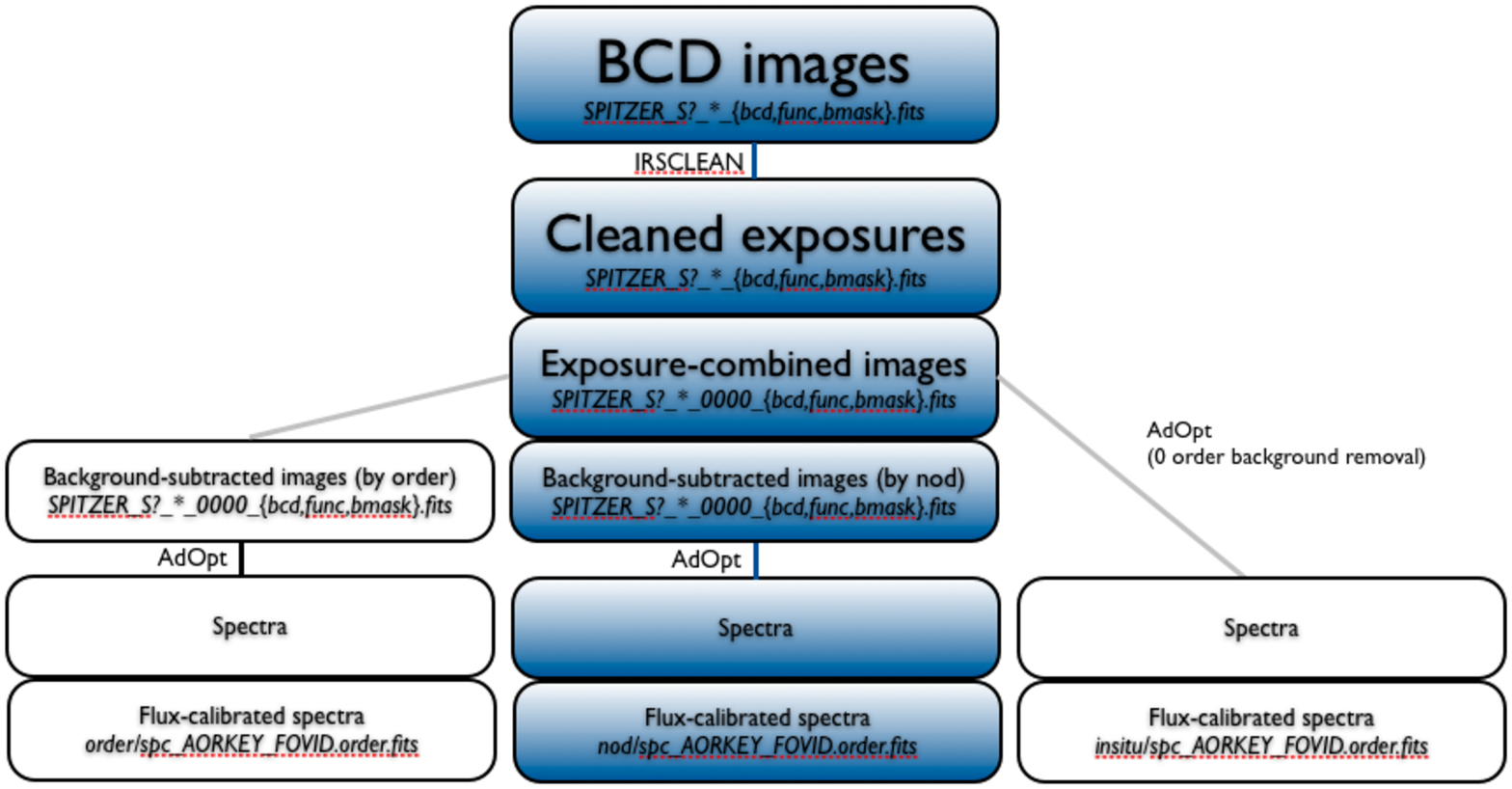}
\figcaption{Basic steps of the extraction pipeline. Only the steps leading to the optimal extraction are shown. Tapered column extraction are performed on the background-subtracted images, and the spectra are then defringed and nod-combined.
The boxes with a dark background correspond to the reference branch which produces the best results in the majority of cases. The naming convention for each available product is shown in the bottom line in each box. \label{fig:pip_part1}}
\end{figure*}

\begin{figure*}
\centering
\includegraphics[angle=90,scale=0.55,clip=true]{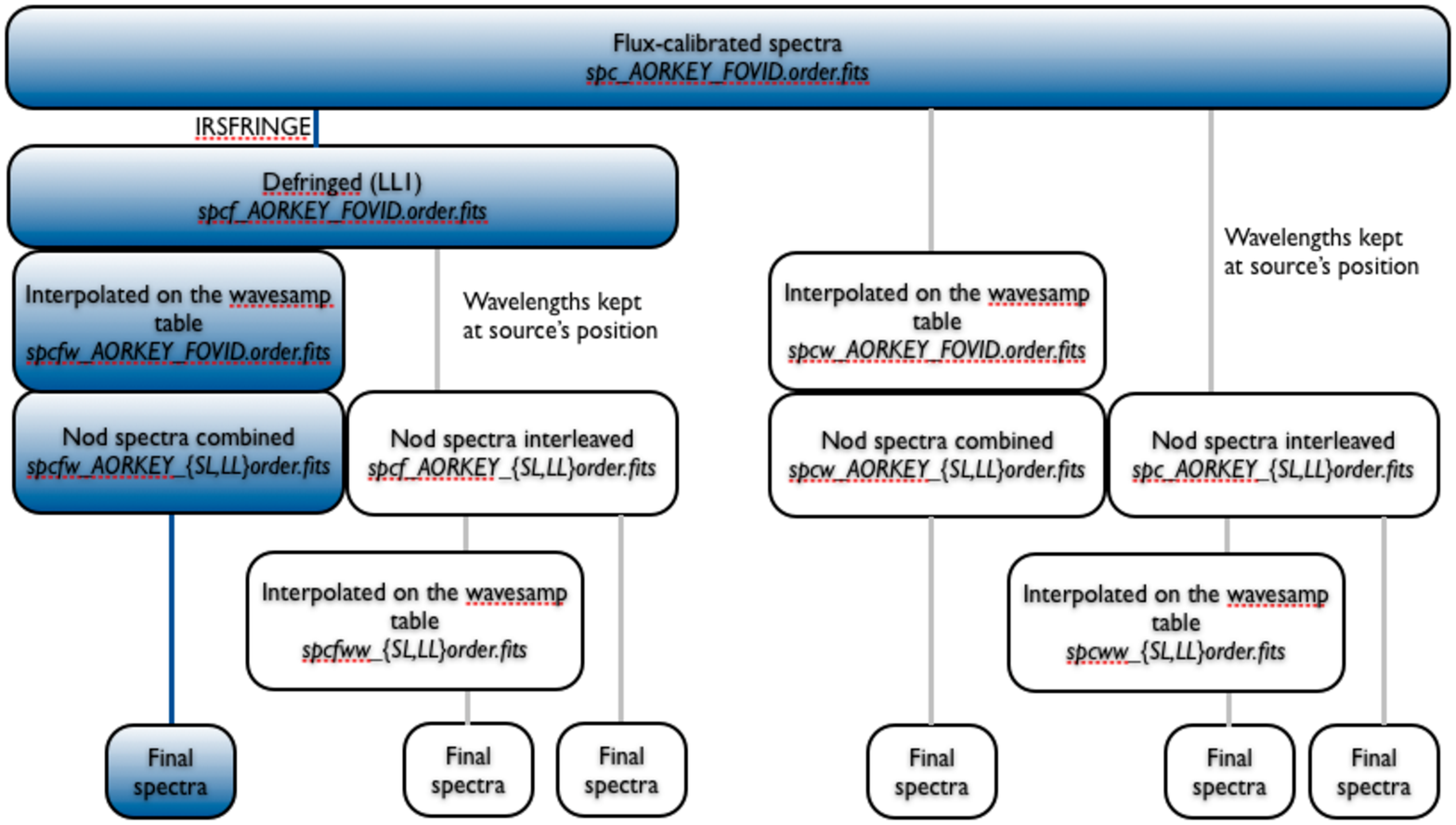}
\figcaption{Basic steps of the extraction pipeline starting from the flux calibrated spectra. See Figure\,\ref{fig:pip_part1} for the figure description. \label{fig:pip_part2}}
\end{figure*}

The CASSIS pipeline provides an alternative to the post-BCD pipeline by enabling a flexible, automatic extraction algorithm that is able to handle very different observations, from barely detected sources to bright sources, from point-like sources to extended sources. CASSIS was built around the SMART-AdOpt tool (Lebouteiller et al.\ 2010) and uses several of its diagnostic tools (e.g., source extent, detection level). The basic steps of the pipeline are shown in Figures\,\ref{fig:pip_part1} and\,\ref{fig:pip_part2} and are reviewed briefly in the next sections; they can be summarized as follows: \\
- \textbf{Steps performed on the calibrated 2D image product}: cleaning of bad pixels, co-addition of individual exposures, removal of background emission, \\
- \textbf{Spectral extraction}: for a given source, two spectra are extracted, one for each nod position.  \\
- \textbf{Steps performed on the 1D spectra}: combination of the nod spectra, flux calibration, defringing.\\


The spectra are gathered in an online database  which provides the single best overall spectrum, combined from both nods, with the maximum S/N.  The database can also be queried  to understand in more detail the various important diagnostic parameters from the pipeline such as the inferred source extent or the detection level  for alternative spectra.

\section{Image pre-processing}


The starting points of the CASSIS pipeline are the BCD images, more exactly the single exposure images also referred to as DCE (Data Collection Events) images. Exposure images are carried through the extraction pipeline along with the corresponding uncertainty image file and the mask of flagged pixels (BMASK, see Sect.\,\ref{sec:cleaning}).

The first public version of CASSIS uses data created with the SSC pipeline release S18.7.0. Internally, this version corresponds to the 4$^{\rm th}$ iteration of the CASSIS pipeline, so that products are labelled ``v4''. Another processing will  eventually be performed using the final S18.18.0 calibration files. New public versions  of CASSIS will be posted as minor modifications in the algorithm warrant a new release, and the online database will be updated accordingly.

\subsection{Bad pixels}\label{sec:cleaning}

The BMASK image contains the flag value of each pixel for a given exposure. CASSIS considers that a pixel is bad if the mask value is greater than 256. Pixels with such flag values have the following issues identified by the SSC BCD pipeline: the flat field could not be applied, the stray-light removal or crosstalk correction could not be applied, pixel is saturated beyond correctable non-linearity in sample(s) along ramp, data missing in downlink in sample(s) along ramp, only one or no usable plane, pixel identified as permanently bad. We refer to the Spitzer/IRS documentation for more details on the flags. Although pixels with BMASK values above 256 could possibly be used, it is important that the pipeline remains conservative in order to minimize potential artifacts. 

Another group of bad pixels, referred to as ``rogue'' pixels, are pixels that misbehave over long periods of time, randomly  changing sensitivity on short timescales. They cannot be properly calibrated. The SSC has released a series of masks identifying the long-term rogues for each IRS campaign. All the bad pixels, including rogue pixels, are cleaned using the IRSCLEAN tool\footnote{The IRSCLEAN package can be found at http://ssc.spitzer.caltech.edu/dataanalysistools/tools.}. The \textit{badfix} method is used with the BMASK file along with a conservative ``super-rogue'' mask that combines rogue pixels from the relevant observation campaign and earlier campaigns. The uncertainty file is cleaned using the same mask as for the data image. We refer to the IRSCLEAN documentation for more details on the cleaning process. 

Note that when several bad pixels are contiguous, the pixel replacement algorithm \textit{badfix} will not fix the pixel(s) in the middle of the cluster. Further pixel verifications are performed in the next CASSIS pipeline steps to attempt a correction.


\subsection{Co-addition of individual exposures}

Before co-adding the individual exposure images (DCEs), CASSIS checks the dispersion of the reconstructed coordinates over the observation duration\footnote{This information is provided by the header keywords SIGRA and SIGDEC populated by the SSC pipeline (Sect.\,\ref{sec:general}).}. If the dispersion is significant, co-adding the images would result in blurring the source spatial profile so that the (optimal) extraction could become unreliable. For this reason, CASSIS co-adds the exposures only when the coordinate dispersion is lower than a certain threshold. 

For a given observation, several cases can be distinguished for the image co-addition depending on the number of exposures available. \\

\textbf{1 exposure.} Single exposures are simply transferred to the next step. \\

\textbf{2 exposures.} Rather than performing a simple combination (such as an error-weighted average), the pipeline takes advantage of having 2 exposures to produce a better result by flagging bad pixels that were not cleaned or could not be cleaned (Section\,\ref{sec:cleaning}). First, pixels are compared using their BMASK value and if one pixel has a higher value than the other, it is ignored. Further bad pixel flagging is achieved by analyzing each column of the image separately. A column can be seen almost as a spectrum, since the wavelength axis is almost parallel to the detector $y$-axis. Outliers are then identified using their deviation from the local flux variance. When the 2 pixels are equally bad, the average is taken. In other cases when one pixel dominates the difference to the local variance, only the pixel from the other image is used. All the other pixels that were not flagged as outliers have error-weighted average fluxes. The final uncertainty on the combined pixel is the sum  of the average of the individual errors and of the flux difference between the 2 pixels to account for the fact that root-mean square (RMS) errors in the input images might be underestimated in some cases. \\

\textbf{More than 2 exposures.} For each pixel, the median over all the images is averaged with the error-weighted-average. This is done to minimize systematic errors due to unreliable uncertainty values. Uncertainties are combined accordingly, i.e., taking the average between the error on the median flux (median absolute deviation) and the error on the average (see also Sect.\,\ref{sec:errors}). 

In a small fraction of the observations  ($\approx0.35$\%) there is a significant dispersion of the pointing coordinates over the exposures (0.05 pixels, i.e., $\approx0.1\arcsec$ in SL and $\approx0.25\arcsec$ in LL) that requires the pipeline to performs steps in a different order:
\begin{itemize}
\item The low-level rogue pixels are removed using the background images (Section\,\ref{sec:lowlevel}). 
\item The individual exposures are extracted separately (Section\,\ref{sec:extraction}).
\item The spectra corresponding to the individual exposures are combined, resulting in one spectrum per module, order, and nod position. The two nod spectra are then combined (Section\,\ref{sec:speccomb}).
\end{itemize}

At this point of the pipeline, there is one image per module, order and nod position. Note that for staring cluster observations, there can be several cycles with several exposures for a given position. In this case, the spectra corresponding to the various cycles are first combined. The online documentation provides specific help for staring cluster observations.

\subsection{Removal of the background emission and of low-level rogue pixels}\label{sec:lowlevel}

Although images were cleaned (Section\,\ref{sec:cleaning}), low-level rogue pixels could remain. It is possible to remove their contribution to the intrinsic source emission by subtracting a background image or a set of images from the same module. This step also allows subtraction of the large-scale background emission (mostly dominated by zodiacal dust emission). Two methods are used:\\

\textbf{Subtraction by nod.} The (single) image corresponding to the other nod position is used for the difference. The SMART-AdOpt algorithm is used to test the presence of a point-like source or partially extended source at the location of the current nod extraction. If the test is positive, another background subtraction has to be used. Note that very extended emission does not modify significantly the source's profile, and it can be removed during extraction (Section\,\ref{sec:extraction}). \\

\textbf{Subtraction by order.} The two nod images corresponding to the other order are (error-weighted) averaged. Contamination at the location of the current nod extraction is also tested.

When the image difference is performed, uncertainties are combined quadratically (see Sect.\,\ref{sec:errors}). In some cases, there are contaminating sources in both the ``\textit{by nod}'' and ``\textit{by order}'' images, so that a third method is used, referred to as ``\textit{in situ}'', which simply removes the extended emission during optimal extraction, i.e., without removing the low-level rogue pixels (Section\,\ref{sec:extraction}).

The final spectrum provided to the user always corresponds to the background subtraction method leading to the best S/N (Sect.\,\ref{sec:diags}). In most case, the best method is the subtraction ``\textit{by nod}'' since it involves a background that is observed shortly before/after the science target (Sect.\,ref{sec:general}). Note that the other version of the spectrum corresponding to the other background subtraction method is available as an optional product.

\section{Pre-extraction diagnostics}\label{sec:preextraction}

\subsection{Spatial extent}\label{sec:extent}

All the images (exposure-combined, background-subtracted) go through the SMART-AdOpt program, which first estimates the source spatial extent. The source extent diagnostic is essential to determine which extraction method is suitable. For point-like sources, the preferred extraction is the optimal extraction which produces the best S/N (Sect.\,\ref{sec:pls_opt}). For partially-extended sources, a ``tapered column'' extraction is preferred since it recovers most of the source's flux.   This extraction includes all flux within the diffraction-limited spatial extent of the spectrum, without weighting by the PSF.  For very extended sources, a full slit extraction is appropriate. 

In order to estimate the spatial extent, AdOpt derives the ratio between the full width at half maximum (FWHM) of the source spatial profile and of the PSF profile. Both profiles are collapsed on the 20 detector rows corresponding to the shortest wavelengths, i.e., where the PSF is the narrowest. The resulting ratio is directly related to the intrinsic spatial extent of the source:
\begin{equation}
S =  P \sqrt{x^2-1},
\end{equation}
where $S$ is the intrinsic FWHM of the source, $P$ is the FWHM of the PSF,  and $x$ is the ratio between the FWHM of the observed profile and the FWHM of the PSF. Note that this method assumes that the source profile can be reproduced by a broadened PSF so it is accurate only for partially extended sources.

\begin{figure}[h]
\centering
\includegraphics[angle=0,scale=0.42,clip=true]{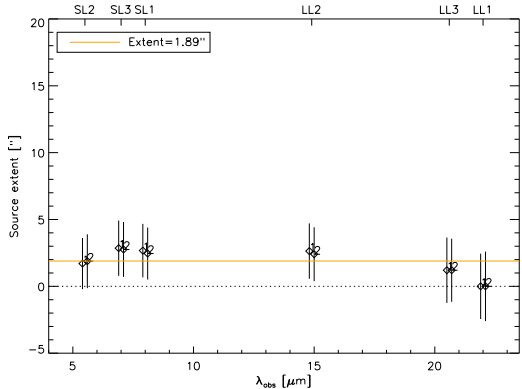}
\figcaption{The intrinsic source extent $S$ is calculated for each module/order spectrum. Two measurements  of $S$ are available for each module and order, one for each nod position. The global extent is calculated using the average value of  $S$ weighted by the detection level in each spectrum.
\label{fig:extent}}
\end{figure}

The global extent of a source is calculated using the average of the $S$ determination for each module and order, assuming the extent is not a function of wavelength. Weights are applied to each value based on the detection level (Section\,\ref{sec:extraction}). Figure\,\ref{fig:extent} gives an example of the output plot produced to compute the spatial extent. 

Determinations of $S$ can be different for each module because the intrinsic source extent might vary with wavelength. Further iterations of the CASSIS processing will include the calculation of the spatial extent for each wavelength element

\subsection{Multiple sources}

Before the extraction of the intended source is performed, the presence of another source in the slit is checked. Depending on the relative brightness between the intended source and the contaminated source, it is possible that the local background emission is not well determined (Sect.\,\ref{sec:extraction}). If there is a positive detection of a contaminating source within the slit, the extraction continues but a flag is carried through for the corresponding module/order spectrum and appears in the result page of the online interface (Sect.\,\ref{sec:interface}). In such cases the users can check the diagnostics plots provided by the interface to judge the quality of the extraction. The spectrum of the serendipitous source is not extracted. 

\section{Spectral extraction}\label{sec:extraction}

Regardless of the source extent, both optimal extraction and tapered column extraction are performed. Hence, for a given source, there are 12 different spectra (2 nod positions, 2 extraction techniques, 3 background-subtraction methods). Depending on the spatial extent (Sect.\,\ref{sec:extent}), and on the best background available (Sect.\,\ref{sec:lowlevel}), the user will be presented with the best extraction choice and the other spectra will remain available as optional products.

\subsection{Optimal extraction}\label{sec:pls_opt}

Optimal extraction uses the PSF profile to weight the pixels from the spatial profile, while tapered column extraction integrates the flux in a spectral window that expands with wavelength. 
Optimal extraction is performed with the AdOpt algorithm which makes use of a super-sampled PSF (Lebouteiller et al.\ 2010). A super-sampled PSF is critical since it can be shifted and resampled anywhere along the slit, making the optimal extraction valid for any source's position. For this reason the AdOpt optimal extraction is an ideal choice for extracting the full set of IRS observations. 

\begin{figure*}
\centering
\includegraphics[angle=0,scale=0.65,clip=true]{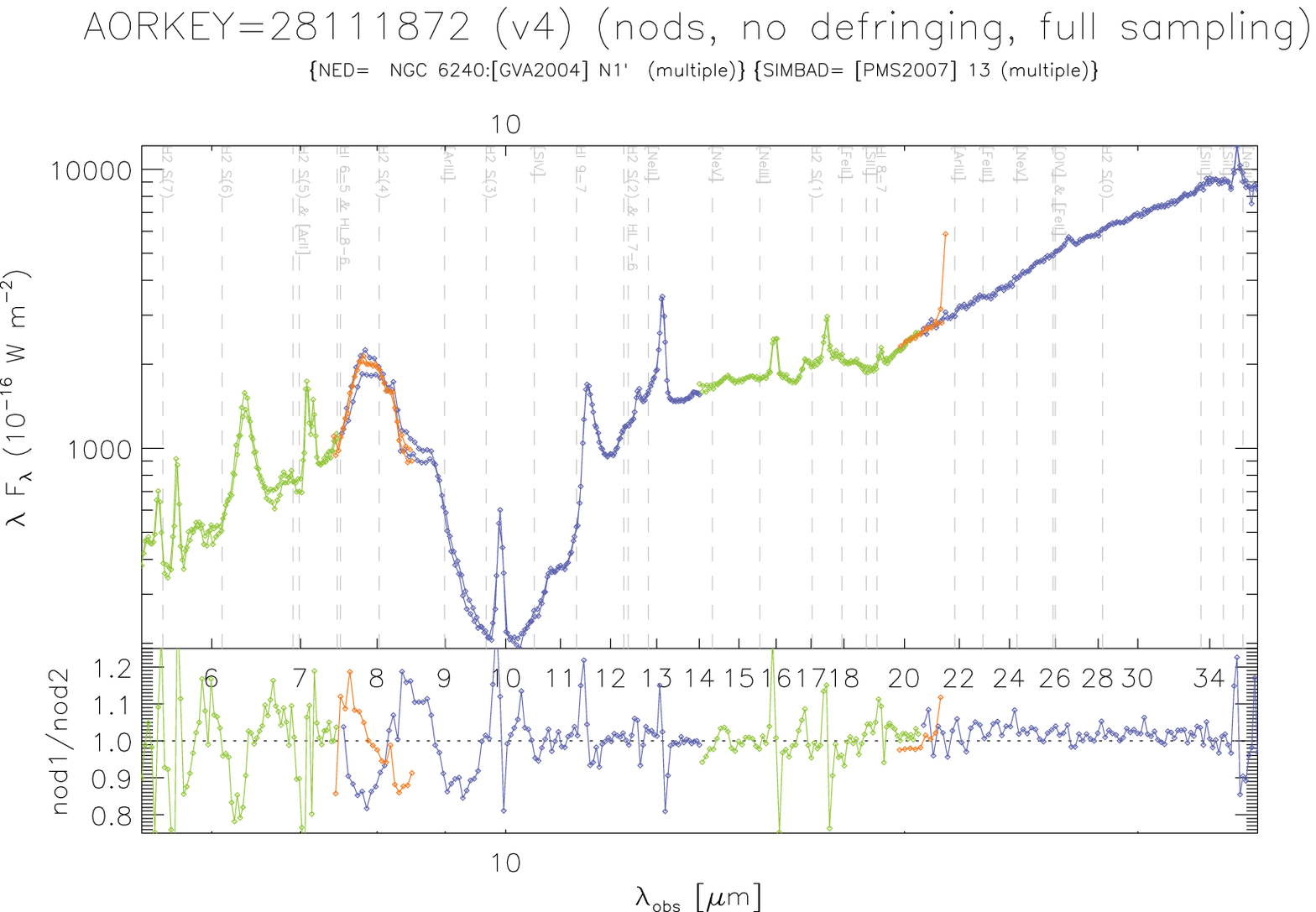}\\
\includegraphics[angle=0,scale=0.6,clip=true]{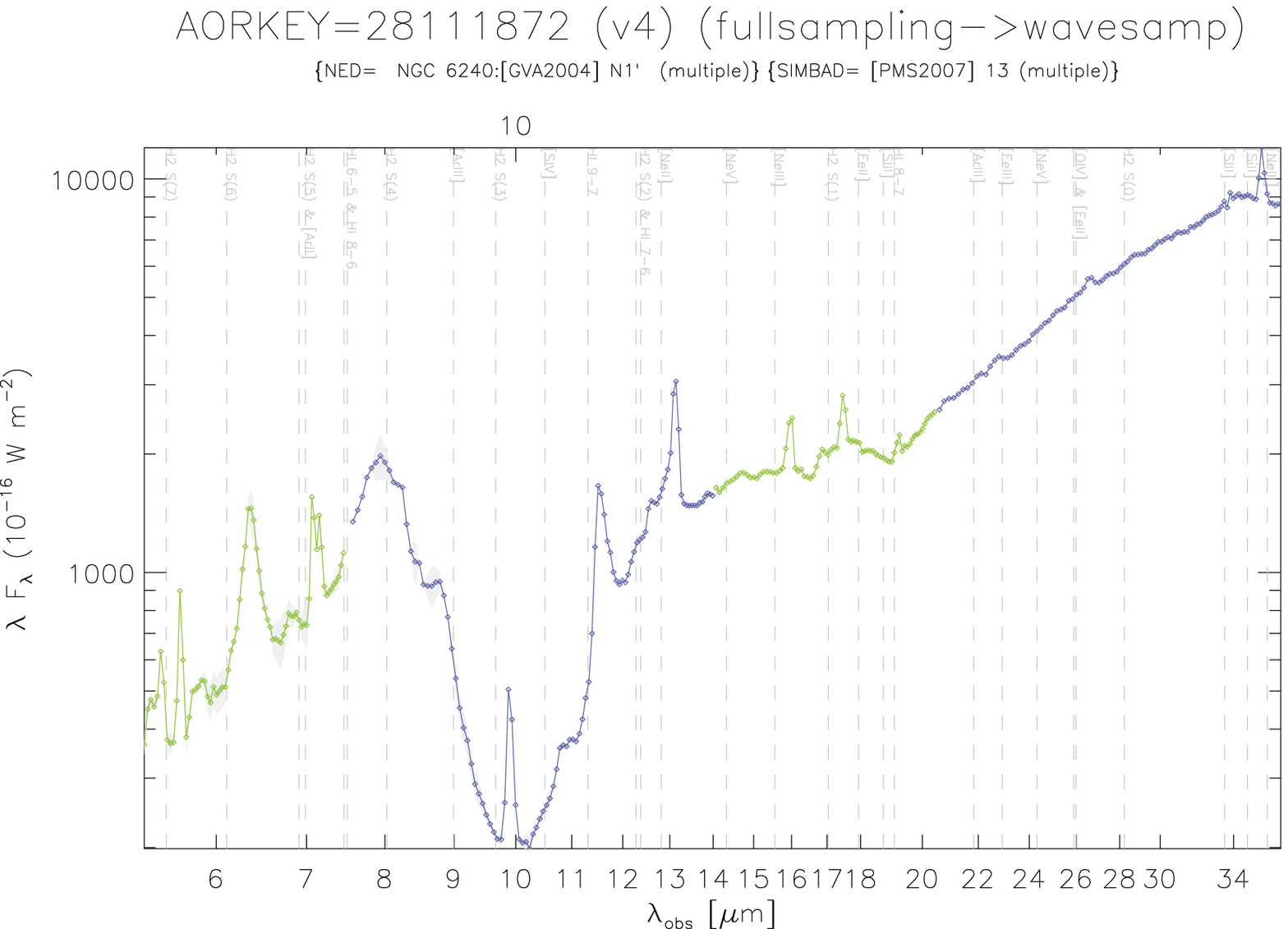}
\figcaption{Example of a CASSIS spectrum for a given source. Only the optimal extraction of the spectrum is shown since the source is point-like. The tapered column extraction always remain available as an optional product. \textit{Top} $-$ The spectrum at each nod position is plotted and the ratio between both spectra is given in the bottom panel. This is an optional plot that the user can choose to inspect for diagnostic purposes. \textit{Bottom} $-$ The two nod spectra have been combined, providing the default product. Vertical dotted lines indicate the position of the brightest atomic and molecular lines. This is the main plot output by the web interface. 
\label{fig:spectra}}
\end{figure*}

Since the CASSIS atlas is initially meant to provide the spectra of  targeted sources, the source finder is limited to positions around a given nod, with a range of $\pm2$\,px (see the pixel size in Table\,\ref{tab:modules}) to account for slight mispointings. The range around the nod position ensures that the right source is extracted even when there is a brighter source in the slit. The source position is then fine-tuned to an accuracy of better than a tenth of a pixel. Examples of spectra are shown in Figure\,\ref{fig:spectra}.

Depending on the geometry, extended emission might still be present despite the background image subtraction (Sect.\,\ref{sec:lowlevel}). This is because the background image does not correspond exactly to the background at the source's position. For this reason, a 0-degree polynomial background is calculated on-the-fly by AdOpt. It is important that other bright sources are not contaminating the observation so that the polynomial background is well estimated (Section\,\ref{sec:preextraction}).


In order to calculate the flux at a given wavelength, SMART-AdOpt performs a multiple linear regression (Lebouteiller et al.\ 2010). 
A quantitative detection level is calculated by the AdOpt algorithm. The detection level represents the percentage improvement between the initial image and the residual image in which the source has been subtracted. For simplicity, the detection level is encoded into an integer value illustrating the quality of the detection: 0 for sources not detected, 1 for tentatively detected, 2 for barely detected, 3 for detected, 4 for well detected. Note that the detection level is accurate only for point-like sources for which AdOpt is currently designed. For extended sources, a significant residual emission can remain after the point-like source extraction, effectively decreasing the detection level parameter.

\subsubsection{Wavelength grid}\label{sec:wgrid}

As explained in Lebouteiller et al.\ (2010), AdOpt extracts detector rows instead of pseudo-rectangles, the latter corresponding to a zone in the image where pixels have the same wavelength. As a result, the output wavelength grid of the extracted spectrum depends on the exact source position. Optionally, the wavelength grid can be interpolated afterwards on a common reference grid (the SSC ``wavesamp'' calibration file). 

The choice of the wavelength grid depends whether spectral resolution or S/N should be privileged. By choosing the observed wavelength grid, the spectral resolution is slightly better than what can be achieved with the optimal extraction of the SSC SPICE software at the expense of S/N. In contrast, the S/N when interpolating the wavelength grid is similar to the optimal extraction in SPICE at the expense of spectral resolution. It must be kept in mind however that the wavelength grid interpolation results in a smoother spectrum but it can make two bad pixels out of one (which is also an undesired effect of pseudo-rectangle extractions). 

Since the choice of the wavelength grid is not only a technical but also a scientific decision, the CASSIS interface always proposes both options to the user. By default however, CASSIS uses the interpolated spectra in order to provide wavelength grids compatible with the tapered column extractions. 


\subsubsection{Nod spectra combination}\label{sec:speccomb}

Although the spectra of each nod are available as optional products, CASSIS produces the nod-combined spectrum, which is the default output. The combination process depends on whether the spectra were interpolated or not (Section\,\ref{sec:wgrid}). We discuss both methods separately. The error treatment is described in Section\,\ref{sec:errors}.

\textbf{Reference ``wavesamp'' wavelength grid (standard product provided as the default)}. The two nod spectra are calculated on the same reference wavelength grid. The co-added spectrum is the error-weighted average of the two nod spectra. The error function (difference between the spectra) is then calculated, smoothed using a multi-resolution algorithm, and used to identify outliers in the individual nod spectra. Pixels are corrected accordingly to their relative discrepancy, i.e., if the pixels in the two nod spectra are outliers, the error-weighted average is used, but if only one pixel is an outlier, the other nod spectrum is used.  

\textbf{Observed wavelength grid (optional product)}. The two nod spectra are first interleaved and aligned (see Figure 4 in Lebouteiller et al.\ 2010). Alignment is performed by calculating the smoothed error function (difference between the spectra). The result is equally split to each nod to align the spectra. Outliers are identified essentially the same way as for the interpolated nod spectra. The only difference is that a pixel identified as an outlier is flagged and given a \textit{not a number} (NaN) value.  
 The final spectrum has twice as many points compared to the individual input nod spectra. It is referred to as the ``fully sampled'' spectrum.

\textbf{Another optional version} of the final spectrum is calculated by interpolating the interleaved spectrum on the reference wavelength grid. Note that this is different from the co-addition of interpolated nod spectra. The interpolation of the interleaved spectrum includes more points and is usually more accurate.

\subsubsection{Flux calibration}

Flux density is converted from e-/sec to Jy using the default option in SMART-AdOpt, i.e., the use of a relative spectral response function (RSRF) derived from the comparison between observations of calibration stars and their theoretical models. The calibration star HR\,6348 was used to produced the CASSIS flux calibration for all modules except LL after campaign 45 (the LL detector bias changed after this campaign). 
To calibrate LL from campaign 45 and on, RSRFs were constructed using
the stars HR~6348 and HD~173511.  A future paper (G.C.~Sloan et al., in preparation) will explain the construction of the templates for these two stars and their photometric calibration.

For point-like sources, the CASSIS interface displays by default the spectrum extracted with optimal extraction\footnote{Using optimal extraction on a partially extended source would result in a slightly underestimated flux density and would slightly modify the spectrum slope, an effect which also exists, though to a lesser extent, for a regular tapered column extraction.}. For partially-extended sources, the CASSIS interface displays the tapered column extraction, which provides a better flux calibration although there is still no satisfactory means of accurately calibrating such sources (Section\,\ref{sec:tapered}). Note that an upgrade of SMART-AdOpt is underway to perform optimal extractions of slightly extended sources. The CASSIS database will eventually include these spectra with the corresponding calibration. 

In about 3\% of the observations the source is not well centered in the slit in the dispersion direction, requiring a special flux calibration. In such cases, a fraction of the PSF lies outside the slit so that the regular optimal extraction fails to fit the proper profile, providing an underestimated flux. Although the manual extraction of AdOpt can solve this problem by modifying on-the-fly the PSF profile as a function of the shift in the dispersion direction, this is not possible automatically. Instead, CASSIS checks the PTGDIFFY header keyword\footnote{The value of this keyword is calculated by the SSC BCD pipeline (Sect.\,\ref{sec:general}).} which gives the pointing error in the dispersion direction between the coordinates requested by the user and the field of view coordinates (effective coordinates at the nod position in the current module and order). This is by no means a definite proof of a genuine offset as the MIR centroid might not coincide with the requested coordinates. In any case, the data is flagged if PTGDIFFY is larger than a given fraction of the slit height and the user is then advised to perform a manual extraction.

\subsection{Tapered column extraction}\label{sec:tapered}

For partially extended sources, the default output spectrum is extracted with a tapered column. The tapered column extraction uses an extraction aperture whose width varies with wavelength because of the varying diffraction limit with wavelength. Depending on the spatial extent derived in (Sect.\,\ref{sec:preextraction}), the extraction aperture width is scaled up to account for all of the source flux.

The flux calibration for tapered column extraction  uses the default SSC flux conversion tables. Note that while the flux might be closer to the truth when using tapered column extraction on a partially extended source as compared to using optimal extraction, there is currently no satisfactory calibration for sources other than point-like sources and very extended sources (significantly larger than the slit). One undesirable effect is that, when using a point-like source flux calibration (whether in optimal extraction or tapered column extraction) on a partially extended source, the LL spectrum contain more flux than the SL spectrum spectra because the LL extraction aperture is larger than the SL aperture. A more advanced version of the flux calibration is currently under investigation and will be included in future CASSIS processings. It uses a theoretical PSF and computes the amount of light lost outside the slit as a function of a broadening parameter.

\section{Defringing}\label{sec:pls_defringe}

Fringes are common in infrared detectors, they produce sinusoidal variations in the spectrum (e.g., Kester et al.\ 2003). For Spitzer/IRS, spectra from the LL1 module are the most affected. The IRSFRINGE\footnote{IRSFRINGE can be found at http://ssc.spitzer.caltech.edu/dataanalysistools/tools.} algorithm is used by CASSIS to remove the fringes in that module only. We limited the defringing solution to two sine functions to prevent an overcorrection of any real structure in the spectrum. Furthermore, since defringing can produce undesired artifacts for sources with a low S/N, only sources with LL1 spectra with S/N$\geqslant$5 are considered.

Note that since defringing is a complex process that uses several hypotheses and parameters, we also provide the uncorrected spectra available as an optional product.

\section{Flux Uncertainties and Calibration}\label{sec:errors}

For all the steps until the spectra combination (Sect.\,\ref{sec:speccomb}), the statistical (RMS) errors  within an individual spectrum are fully propagated using the standard equation: 
\begin{equation}
s = \frac{ \sqrt{ \sum_{i=0}^N w_i^2 \sigma_i^2 } } { \sum_{i=0}^N w_i },
\end{equation}
where $w$ is the weight and $\sigma$ is the uncertainty for the image (or spectra) number $i$. The weighting factor, although arbitrary, depends on the error distribution. The error distribution is mostly poissonian (photon count noise) for sources with fluxes significantly above the detector readout noise, while it becomes ``normal'' for very faint sources. The pipeline processing steps then naturally tend to ``normalize'' errors. We chose to consider weights as $1/\sigma^2$ throughout the pipeline. 

The combination of the spectra introduces  a systematic error illustrated by the flux density differences between the two nod spectra. This difference is usually due to the presence of other sources in the slit  affecting the background subtraction, or to pointing errors resulting in a slight shift in the dispersion direction. The two errors (RMS and systematic) are given as two different fields in the output files.

Another source of systematic error is the flux calibration itself (G.C.\ Sloan et al.\, in preparation). The corresponding uncertainty depends on the stellar template(s) used  (i.e. on the uncertainties inherent to the stellar models themselves) and on the dispersion in the various spectra of the calibration stars used to build the RSRF. With the current flux calibration in CASSIS, we estimate this systematic error to be less than $\approx2$\%. Wavelength ranges near the edges of detectors are not so well calibrated because of instrumental effects such as light leaks.  Such uncertainties are not yet estimated for the first public CASSIS version (``version 4'').

\section{Products and diagnostics}\label{sec:diags}

The main and final product is the calibrated spectrum resulting from merging the spectra of the various modules, nods, and spectral orders. The default spectrum presented to the user depends whether the source is point-like (optimal extraction was selected) or partially-extended (tapered column extraction was selected). The default spectrum includes  defringing (Sect.\,\ref{sec:pls_defringe}), and the wavelength grid is the reference ``wavesamp'' (Sect.\,\ref{sec:wgrid}). 
The background subtraction method used for a given module and order (Sect.\,\ref{sec:lowlevel}) is the background providing the best detection level according to the AdOpt algorithm. In less than 2\%\ of the observations, the presence of a contaminating source in the background image forces the default spectrum to be presented without any background subtraction. 

The other (non default) versions of the spectra are also available as optional products (optimal or tapered extraction, specific background subtraction method, specific wavelength grid, no defringing, etc.). These versions are  intended for advanced users who wish to inspect in detail the spectra and to compare the various extraction techniques  or even the various wavelength scales available. The online documentation and  interface advises  users to check systematically the optional products especially in the first year of CASSIS since improvements might be needed early on. The various products and the naming convention are the ones shown in Figures\,\ref{fig:pip_part1} and \ref{fig:pip_part2}. 

 The most important parameters defining the output diagnostics for a given AORkey are  the \textbf{coordinates}, reconstructed from the source's position inferred for each spectral extraction with AdOpt; the \textbf{detection level}, corresponding to the maximum detection level in the spectra of each nod, module, and order; and 
the \textbf{spatial extent}, described in Sect.\,\ref{sec:extent}.

In the final step of the pipeline, the extracted source is resolved using the Nasa/IPAC Extragalactic Database (NED\footnote{http://nedwww.ipac.caltech.edu/}) and the SIMBAD Astronomical Database\footnote{http://simbad.u-strasbg.fr/simbad/}. 
In cases where multiple sources fall within the PSF beam, the closest matching source is chosen and a special flag is given. The spectroscopic redshift is inferred using NED and the coordinates of the resolved source. No redshift is returned if multiple sources are found within the PSF beam.  The source identification and redshift are provided to the user and are part of the CASSIS atlas. Photometric and spectroscopic redshifts from VIZIER\footnote{http://vizier.u-strasbg.fr/viz-bin/VizieR} catalogs will be included in future CASSIS versions.

\section{Online interface}\label{sec:interface}

The online CASSIS interface offers basic functionalities such as querying by AORkey, program ID, or coordinates.  Eventually, additional parameters will be added to allow querying in various other ways, such as by source spatial extent or spectral properties, or the redshift (Sect.\,\ref{sec:diags}). 

The results page includes observational parameters, extraction diagnostics, and the links to download spectra. The interface also enables viewing and analyzing the spectra with Virtual Observatory (VO) tools. 
We currently interoperate with the International Virtual
Observatory Alliance at two levels: we permit downloading  of
spectra in the IVOA Spectral Data Model\footnote{http://www.ivoa.net/Documents/SpectrumDM/index.html}, so that they
may be processed with VO spectral analysis tools such
as VOSpec and SPLAT.  We also allow running the
VO-enabled SPLAT applet directly on a given dataset.  Eventually, we plan to support Simple Spectral Access
Protocol (SSAP\footnote{http://www.ivoa.net/Documents/latest/SSA.html}) discovery through Virtual observatory
search tools.

\section{CASSIS Source Statistics}

\begin{table*}
\begin{center}
  \caption{Number of observations and distinct sources in the CASSIS atlas per scientific category.}
  \label{tab:stats}
  \begin{tabular}{l l l l}
  \hline
 Category & Observations\tablenotemark{a} & Detections\tablenotemark{b} & Distinct sources\tablenotemark{c}  \\
  \hline
 High-z galaxies & 1252 & 504 & 503 \\
 Intermediate-z galaxies & 847 & 753 & 753 \\
 Nearby galaxies & 484 & 289 & 286 \\
  Local Group galaxies & 561 & 546 & 542 \\
  AGN/quasars/radio-galaxies & 1535 & 1424 & 1385  \\
 Interacting/mergers & 156 & 147 & 147 \\
 ULIRGS/LIRGS & 631 & 601 & 601 \\
 Starburst galaxies & 123 & 120 & 120 \\
 Galaxy clusters & 57 & 52 & 51 \\
 Cosmology & 3 & 1 & 1 \\
 Extragalactic jets &  5 & 5 & 2 \\
  Gamma-ray bursts & 5 & 5 & 5 \\
  Compact objects & 57 & 56 & 53 \\
Galactic structures & 13 & 12 & 12 \\
 \hline    
Massive stars & 155 & 118 & 118 \\
Evolved stars & 1256 & 1164 & 1074 \\
Brown dwarfs &  287  &  282  &  274  \\ 
 Stellar population & 218 & 218 & 164 \\
 \hline    
  Star formation & 577 & 548 & 526 \\
  Young stellar objects & 1670 & 1556 & 1492 \\
Circumstellar disks & 2464 & 2417 & 2379 \\
  Extra-solar planets &  7 & 7 & 4 \\
 \hline    
  ISM & 884 & 768 & 768 \\
  HII regions &  57 & 44 & 44 \\
  \hline
  \hline
  Total & 13264 & 11637 & 11304  \\
  \hline
  \end{tabular}
  \tablecomments{The following categories currently have no spectra in CASSIS: satellites, comets, planets, kuiper-belt objects, near-earth objects, asteroids, star clusters, zodiacal dust, and dark matter.}
  \tablenotetext{a}{The number of observations includes the number of AORkeys and the number of pointings for the cluster observations.}
\tablenotetext{b}{Detection level greater than 1, i.e., sources at least barely detected. The difference between the number of observations and the number of detections is often due to non-detections in background observations designed as separate AORkeys.}
\tablenotetext{c}{Distinct sources are found by matching observations within $4\arcsec$ of each other.}
\end{center}
\end{table*}

\begin{figure*}
\includegraphics[angle=0,scale=0.65,clip=true]{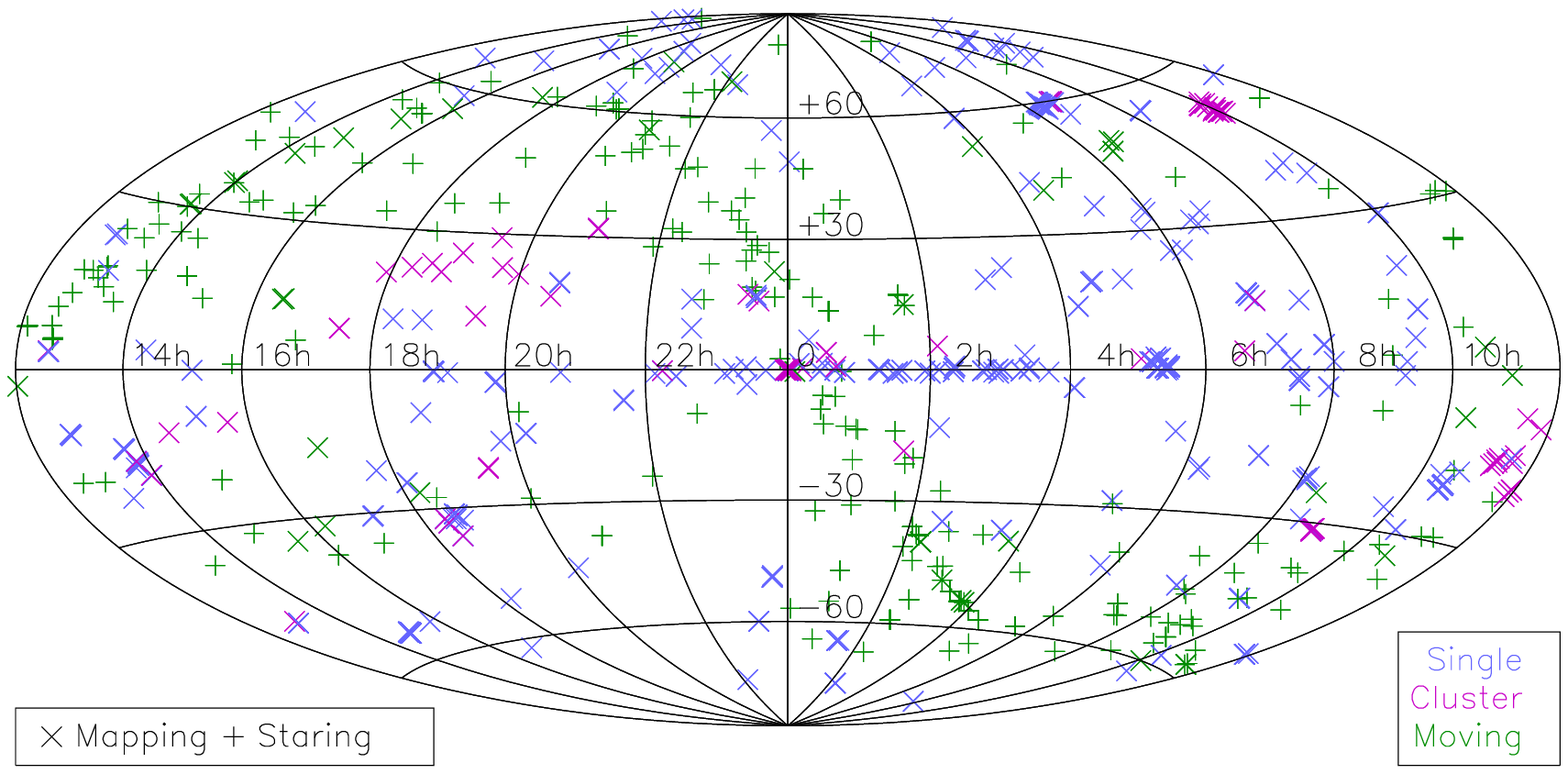}\\
\includegraphics[angle=0,scale=0.65,clip=true]{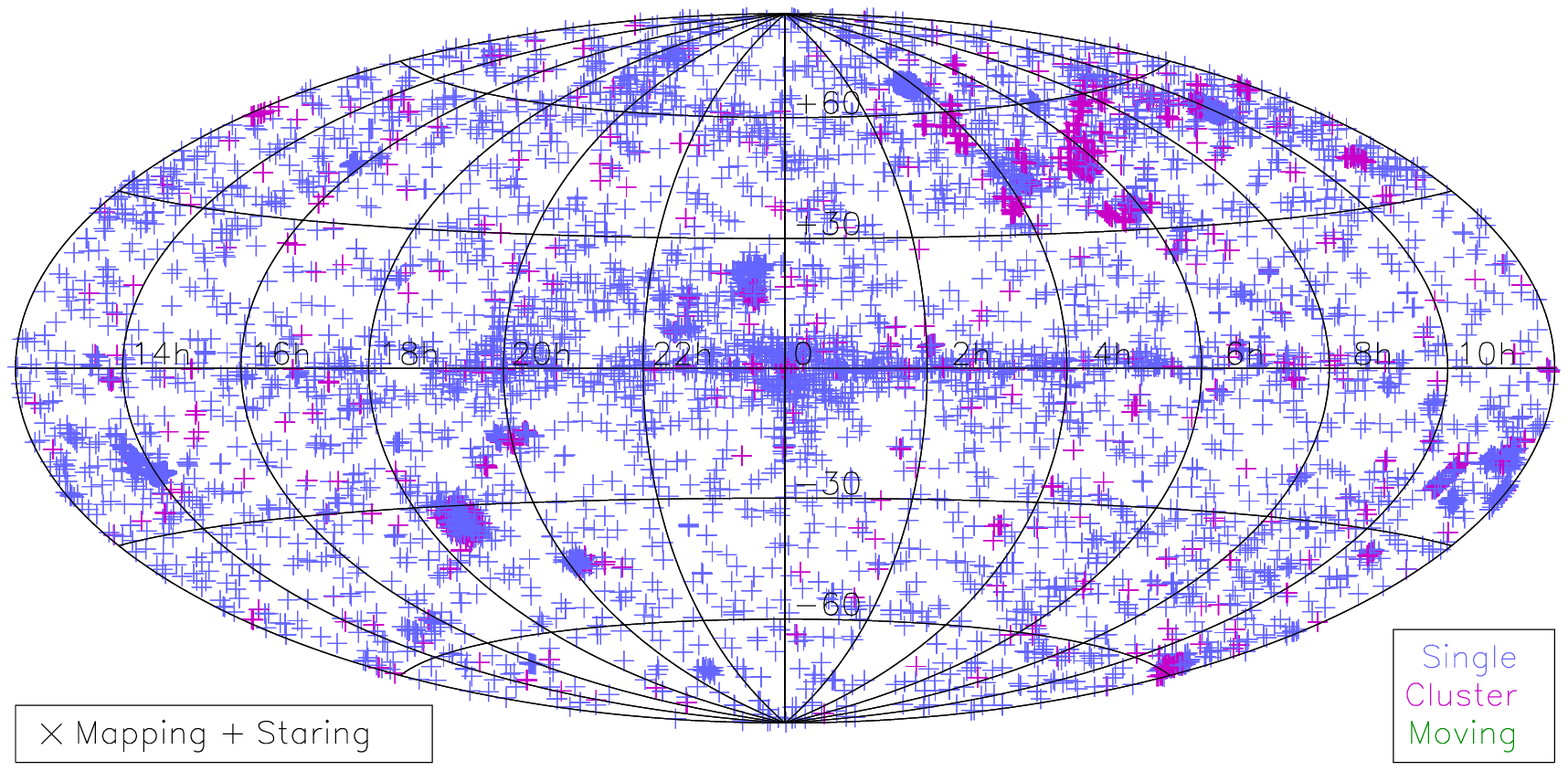}\\
\includegraphics[angle=0,scale=0.65,clip=true]{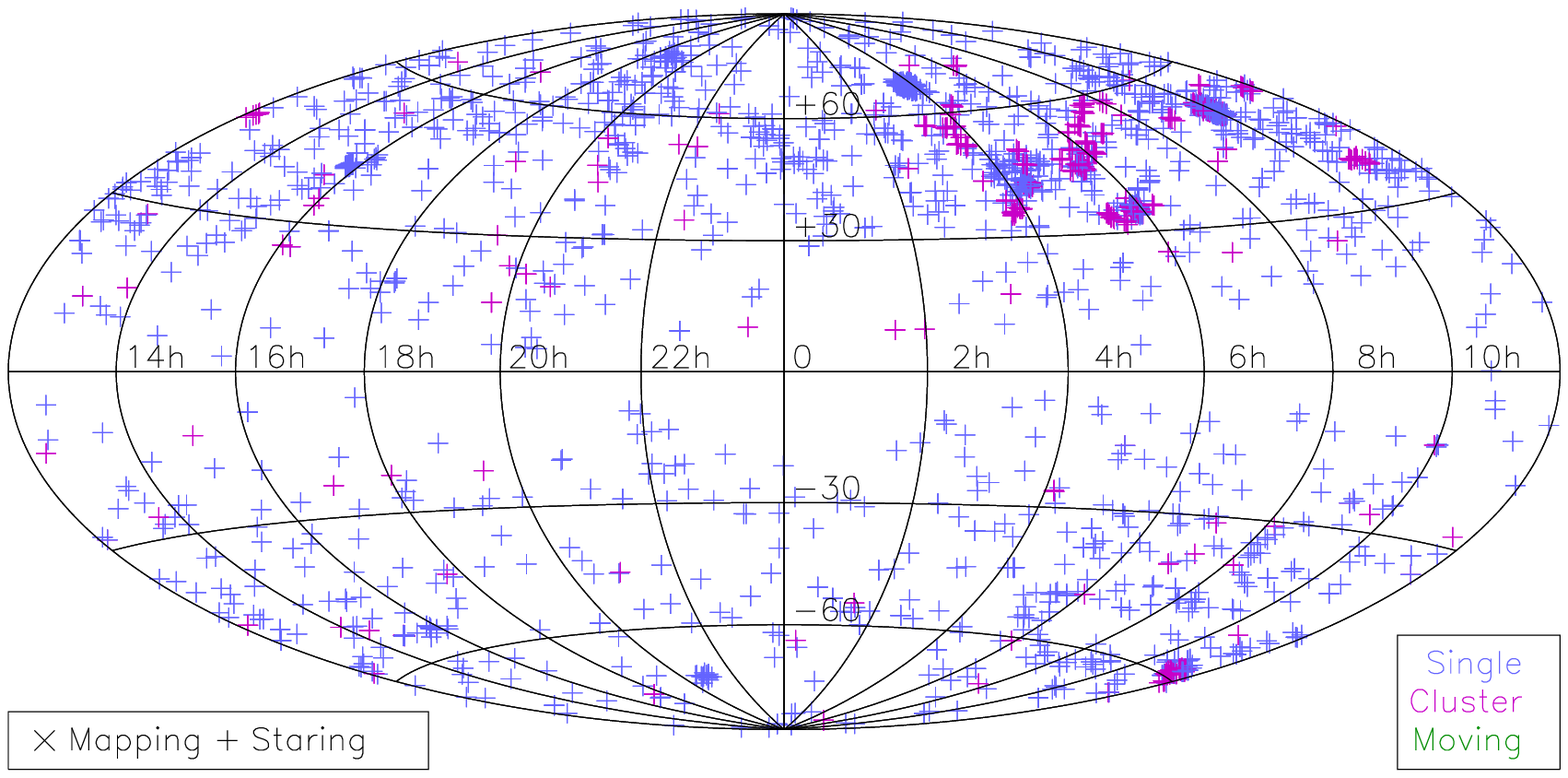}
\caption{IRS observations projected on the sky. The top figure shows the mapping observations and the moving objects which are not included in the atlas. The middle figure shows the observations with spectra extracted in CASSIS, corresponding to staring observations (single or cluster). The bottom figure shows the subset of sources from CASSIS with associated redshifts. \label{fig:irsobs}}
\end{figure*}

The CASSIS atlas currently includes 13\,264 different observations, i.e., most of the low-resolution observations ever performed in the staring mode (single and cluster mode) by the IRS instrument. Figure\,\ref{fig:irsobs} shows the distribution of observations in the sky. Among the CASSIS sources, about $90\%$ are at least barely detected. According to the detection parameter from AdOpt, we find that $\approx73$\% of the objects are well detected (level 4, see Sect\,\ref{sec:pls_opt}), $\approx5$\% are detected (level 3), $\approx7$\% are faint (level 2), $\approx4$\% are barely detected (level 1), and $\approx10$\% are not detected (level 0).

\begin{figure}
\centering
\includegraphics[angle=0,scale=0.42,clip=true]{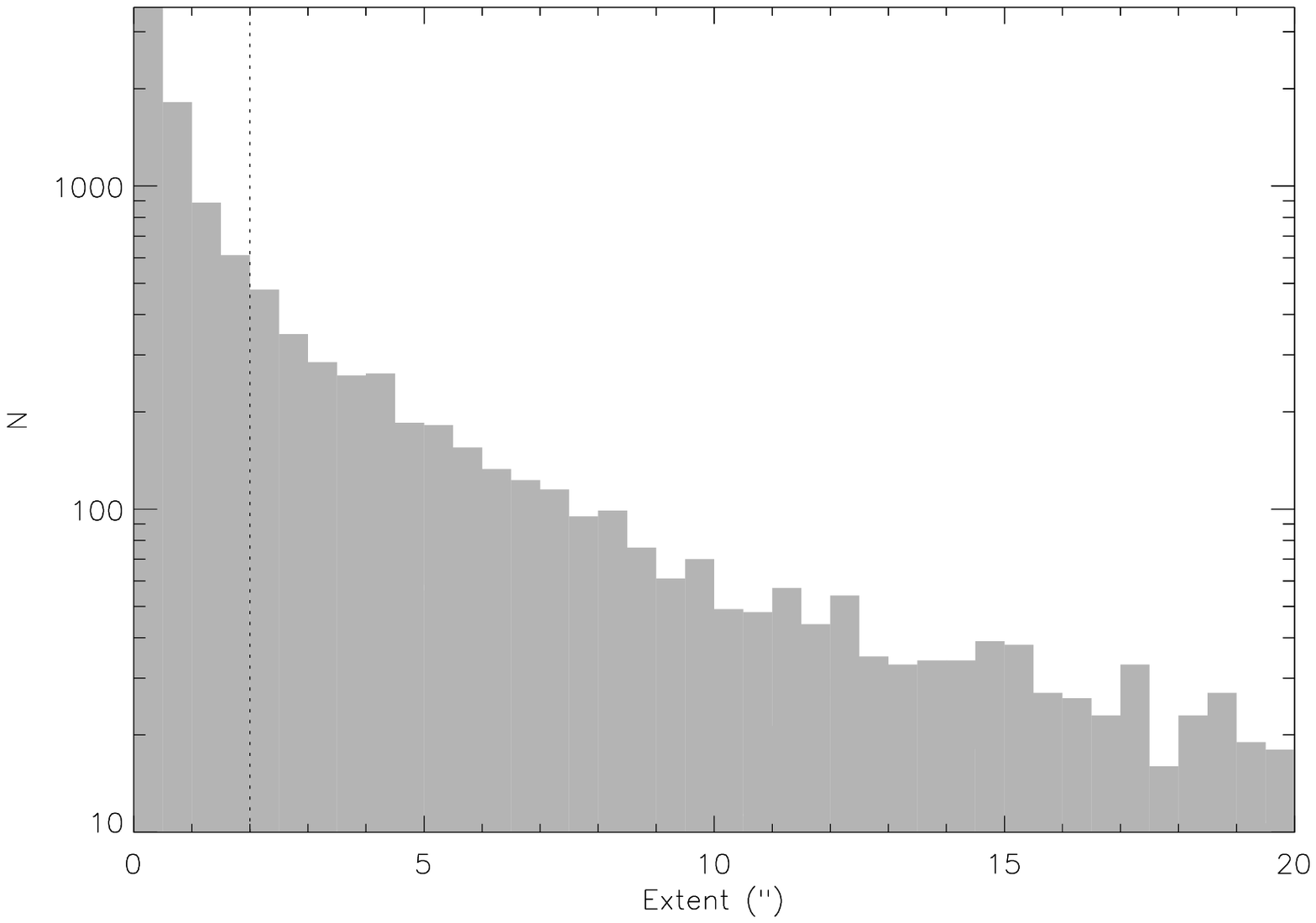}\\
\includegraphics[angle=-90,scale=0.26,clip=true]{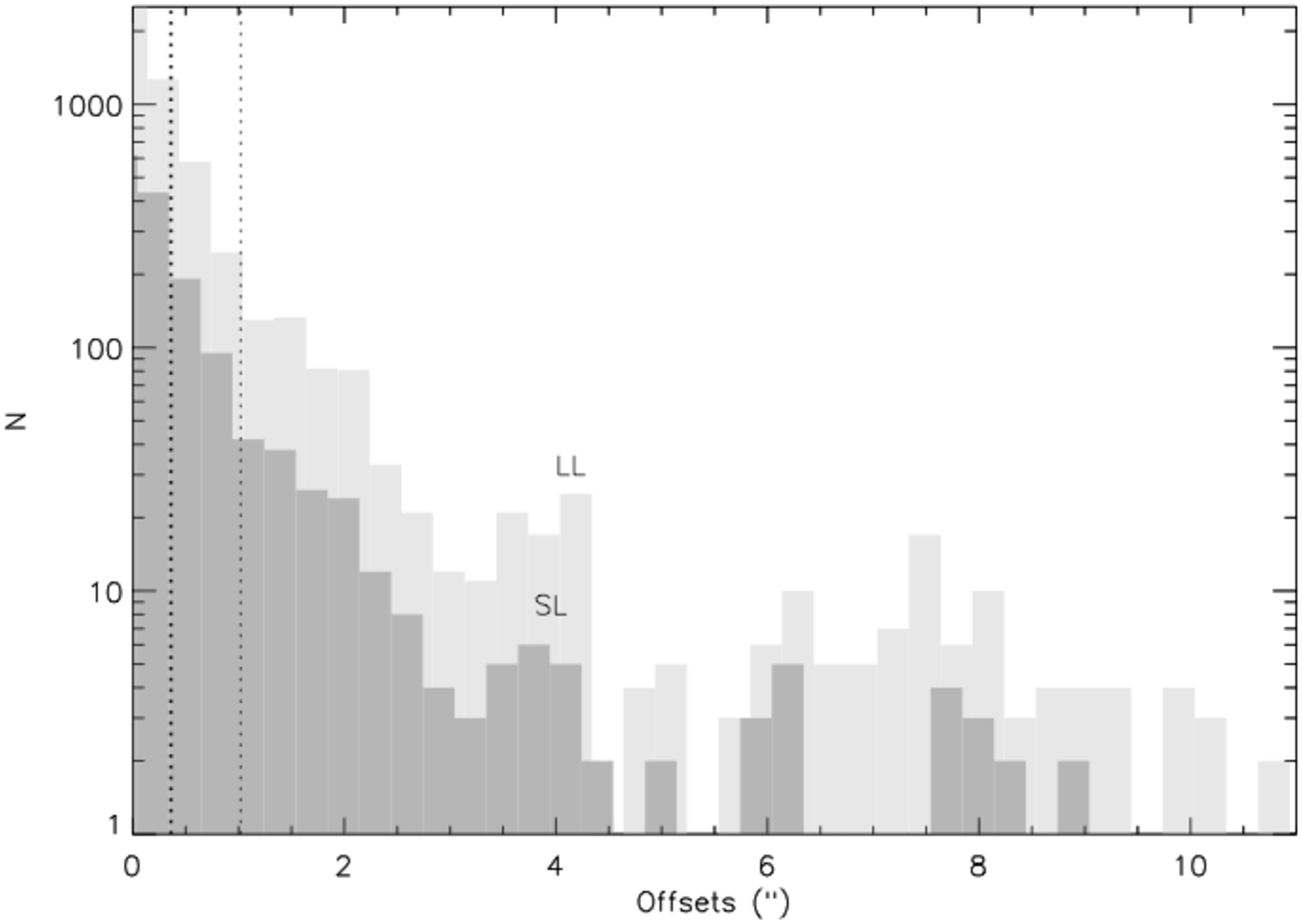}
\figcaption{\textit{Top} $-$ Spatial extent of sources with a level detection above 2, i.e., sources at least barely detected (see Sect.\,\ref{sec:pls_opt} for the determination of the detection level). The vertical dotted line corresponds to an extent of $2\arcsec$ below which sources can be safely considered as point-like. \textit{Bottom} $-$ Pointing offset distribution (in the cross-dispersion direction) of sources with a level detection above 2, i.e., sources at least barely detected. The 2 vertical dotted lines indicate a fraction of 0.2 pixels in SL ($0.36\arcsec$) and in LL ($1.2\arcsec$).
\label{fig:histoextent}}
\end{figure}

Figure\,\ref{fig:histoextent}a shows the histogram of the spatial extent distribution. Most sources (70\%) have an inferred intrinsic extent below $2\arcsec$ and can be considered point-like (see Sect.\,\ref{sec:extent}). Figure\,\ref{fig:histoextent}b shows the histogram of the pointing offsets, i.e., the source position in the cross-dispersion direction of the slit. Most sources are found within a fraction of a pixel from the exact nod position.

\begin{figure}
\centering
\includegraphics[angle=0,scale=0.42,clip=true]{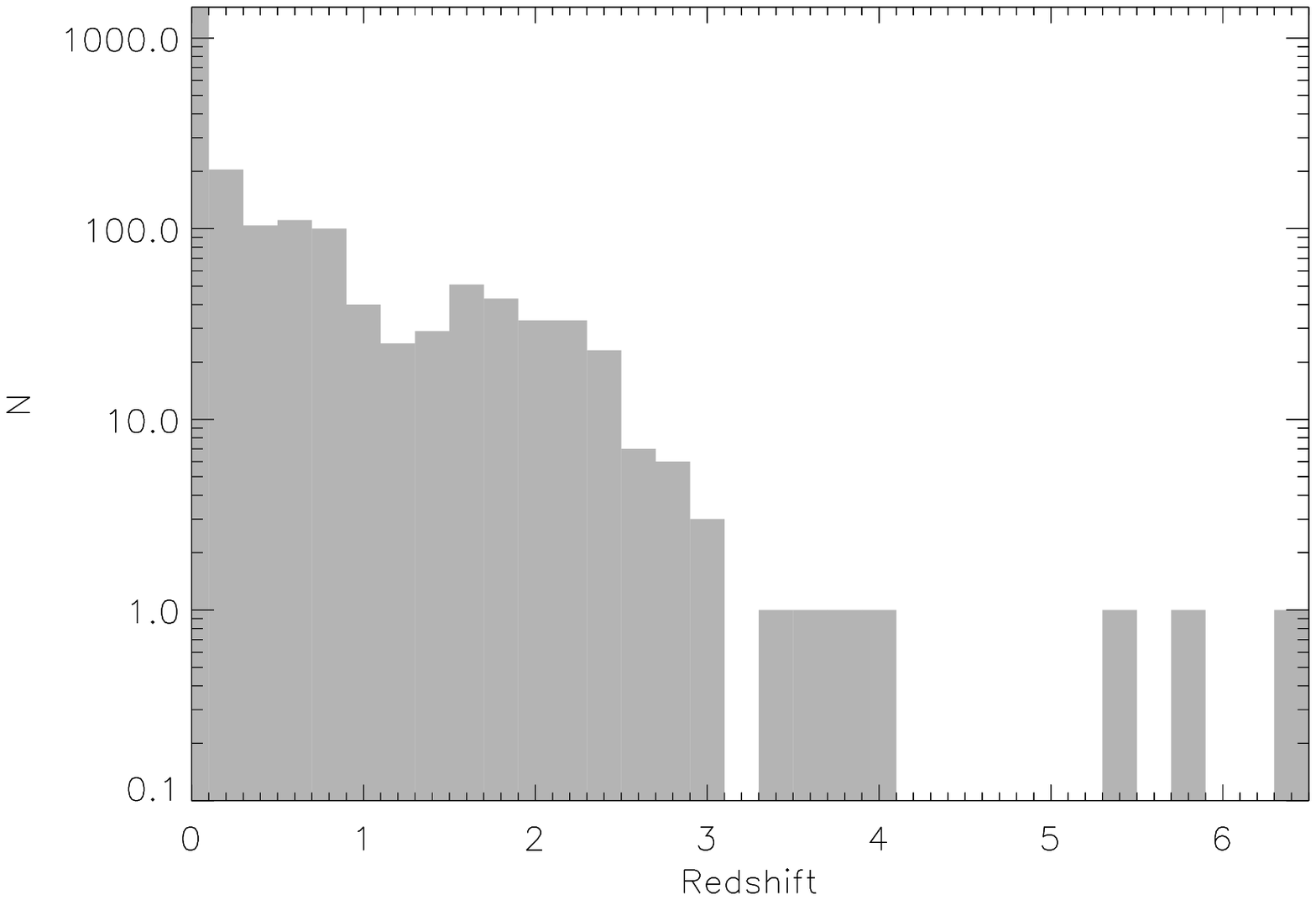}
\figcaption{Redshift distribution of sources with a level detection above 2, i.e., sources at least barely detected (see Sect.\,\ref{sec:pls_opt} for the determination of the detection level). 
\label{fig:histoz}}
\end{figure}

\begin{figure*}
\includegraphics[angle=0,scale=0.85,clip=false]{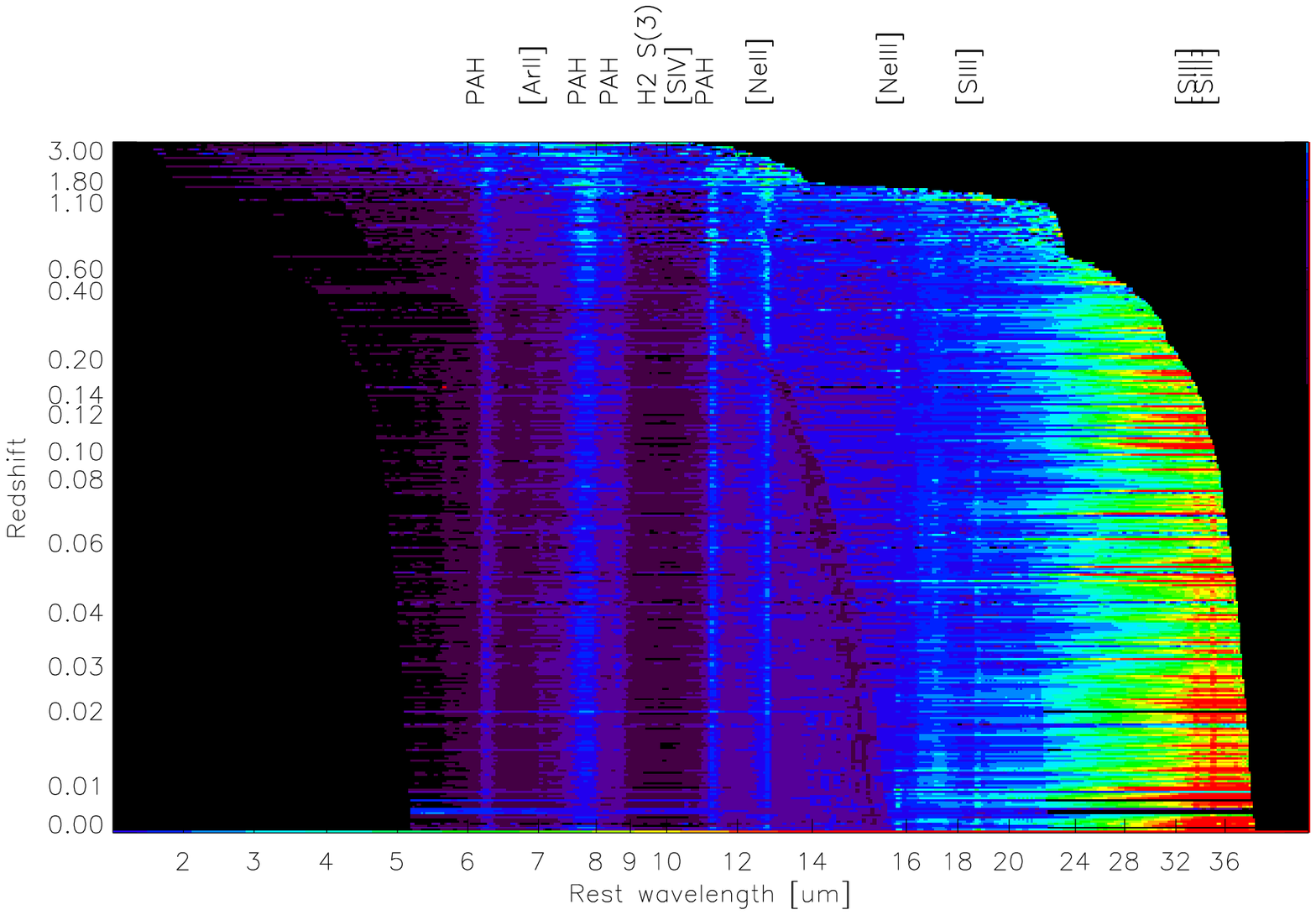}
\caption{Rest-frame spectra of extragalactic objects are plotted as rows, the light scale scaling with the flux density. Only sources with significant PAH emission are included in this sub-sample. The redshift was associated with NED from the coordinates of the extracted source. A few black curves can be seen that corresponds to noisy regions in each module.
\label{fig:stack}}
\end{figure*}

For about 2\,118 distinct sources, a redshift could be associated using NED and the spatial coordinates of the extracted source (see Sect.\,\ref{sec:diags}). The redshift distribution of the detected sources is shown in Figure\,\ref{fig:histoz}. We selected the sources with significant PAH emission from this sample to build the stack of Figure\,\ref{fig:stack}. Spectra are plotted in the rest-frame. The good alignment of the known spectral features indicates that the redshift association provides reliable values. Note that the sources in this example span a wide variety of objects, the only common property being a spectroscopic redshift determination.

\section{``Build your own'' samples}\label{sec:build}

\begin{figure*}
\includegraphics[angle=0,scale=0.5,clip=true]{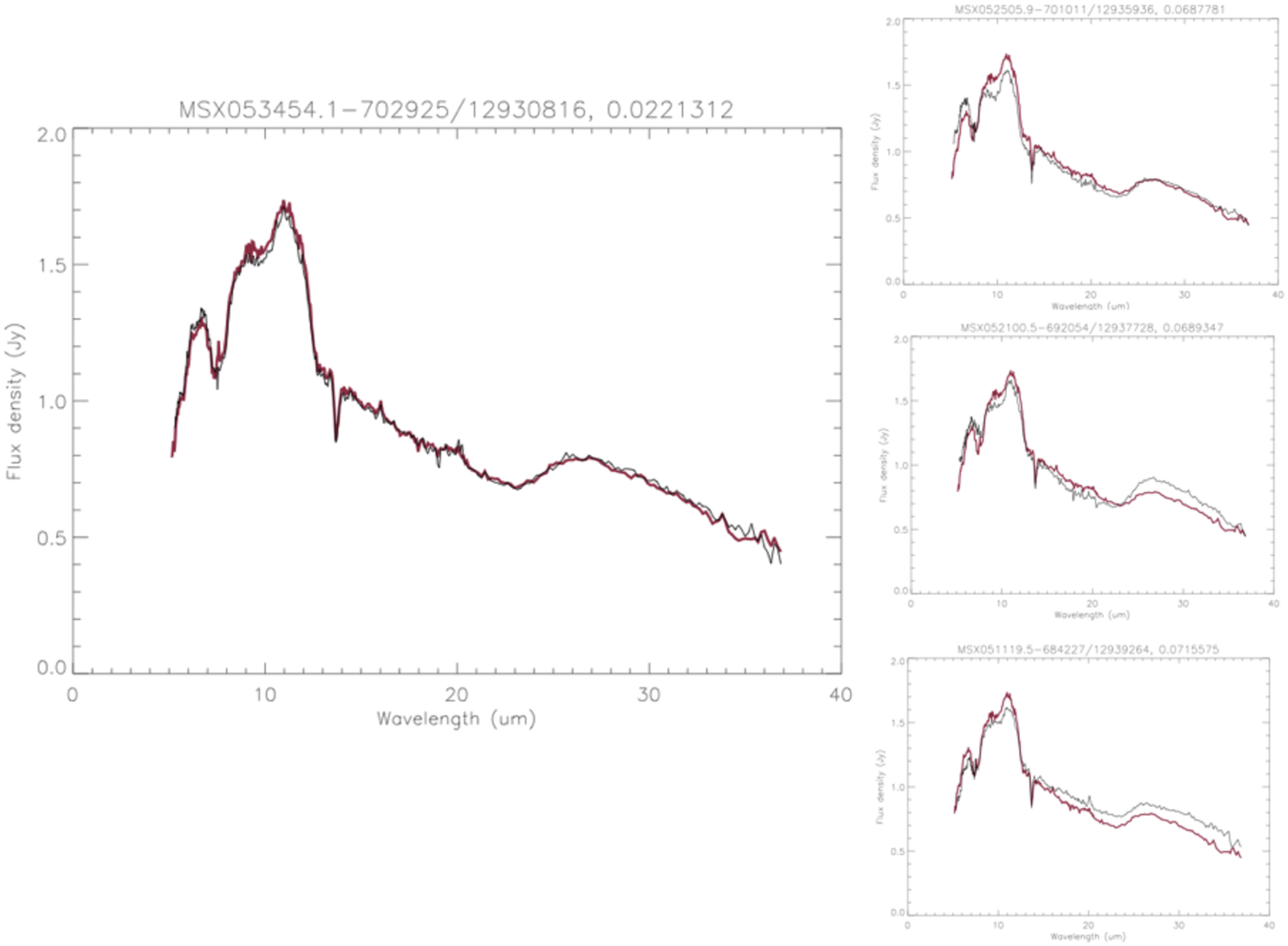}
\caption{The red spectrum (corresponding to a star) was given as an input to find matching spectra in the database. The criterion in this case was to match the spectra over the whole wavelength range. The first hit (\textit{left}) corresponds to the input object itself. The few first other hits are shown on the right. The match can be performed over any wavelength range (observed or rest-frame).
\label{fig:match}}
\end{figure*}

A specific aspect of CASSIS is that the flux density at every wavelength element of the reference (wavesamp) table is contained in the database. Therefore, spectra themselves can be used in a query. For instance, it is possible to query the database for spectra matching a given reference spectrum (model, user-defined, or existing spectrum in the database).  The match can be performed on the full wavelength scale or on any sub-region, for example around a given spectral feature (see example of Figure\,\ref{fig:match}). 

Furthermore, it is possible to perform operations on the spectra to satisfy a given query. For instance, a few simple examples of semantic searches could be: \\
- \textit{``select sources where $f(14, 0.2)>f(5, 0.2)$''}, for spectra with a positive slope from 5\mic\ to 14\mic. The continuum flux is calculated by integrating a $0.2$\mic\ window around a central wavelength, \\
- \textit{``select sources where $f(6.2, 0.2)-0.5\times [f(5.8, 0.2)+f(6.4, 0.2)]>0.1$''}, for spectra with significant emission in the 6.2\mic\ PAH band. \\
- \textit{``select sources where ${\rm SNR}(10, 1)>5$''}, for spectra in which the S/N is greater than 5 between 9 and 11\mic. \\
- \textit{``select sources where (extent$<2$ and globaldetlvl$=4$ and $z<0.1$)''}, for observations of point-like sources with a detection level equal to 4 (well detected) and with a redshift below 0.1. \\

More sophisticated measurements (e.g., involving line fits) for the full CASSIS sample will be provided in a later step. The sample of extragalactic sources with known redshifts is currently being investigated, including the 6.2\mic\ PAH and the silicate strength values (H. Spoon et al., in preparation). The results will be eventually incorporated in the CASSIS database and will be usable by queries.  The mid-infrared classification from Spoon et al.\ (2007), using these 2 spectral parameters will be also included in the database. 

\begin{figure*}
\includegraphics[angle=0,scale=0.5,clip=true]{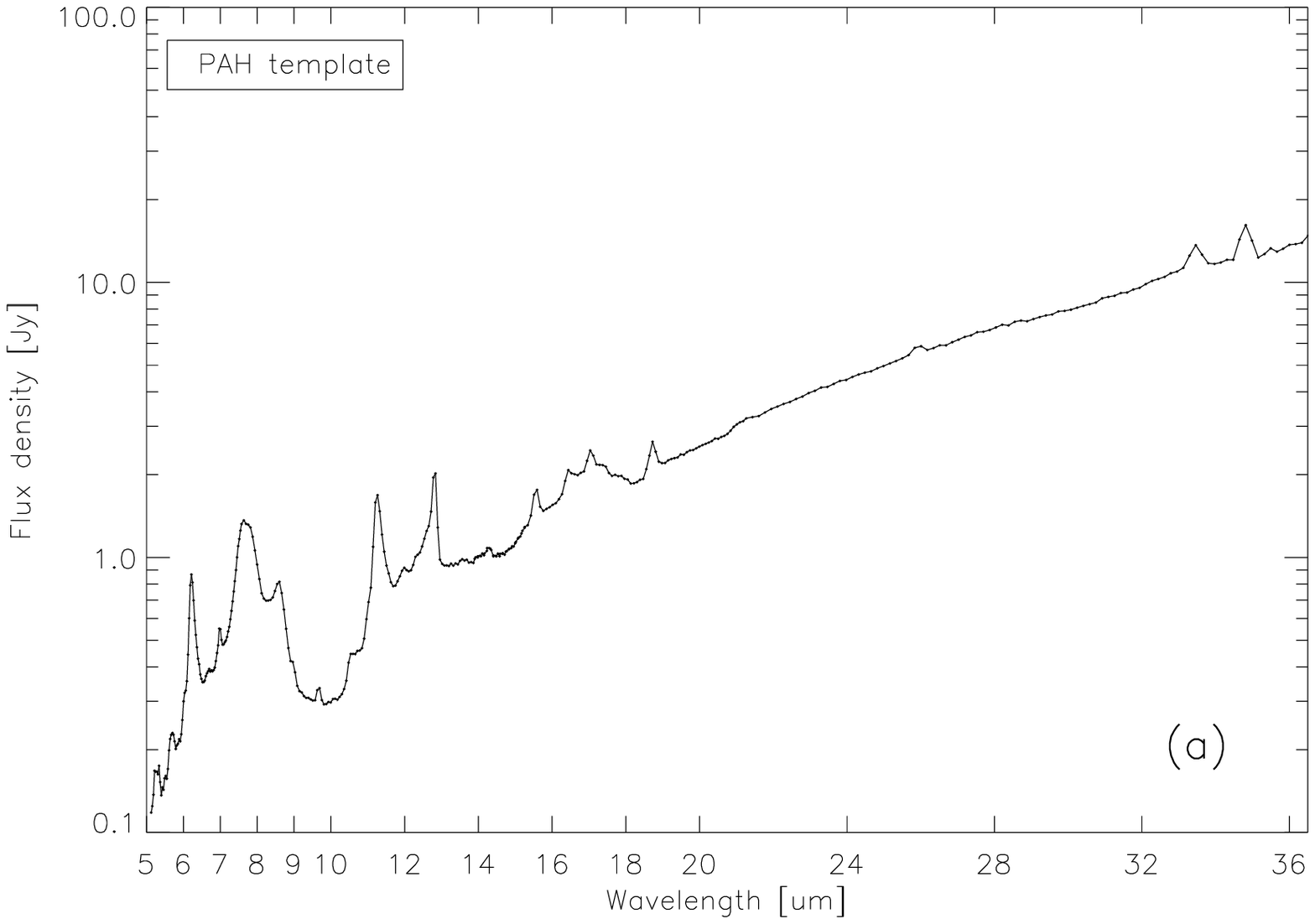}
\includegraphics[angle=0,scale=0.5,clip=true]{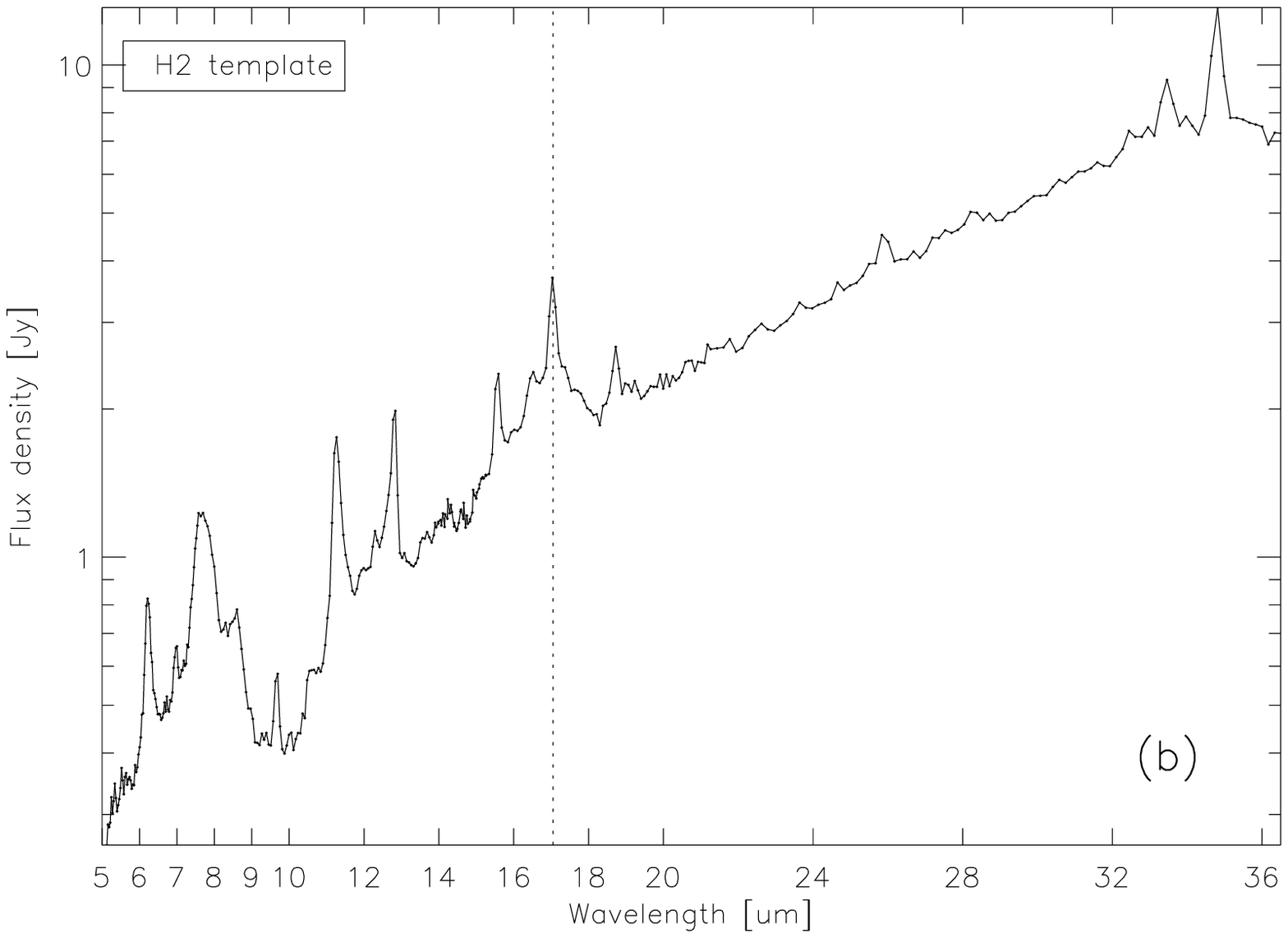}\\
\includegraphics[angle=0,scale=0.5,clip=true]{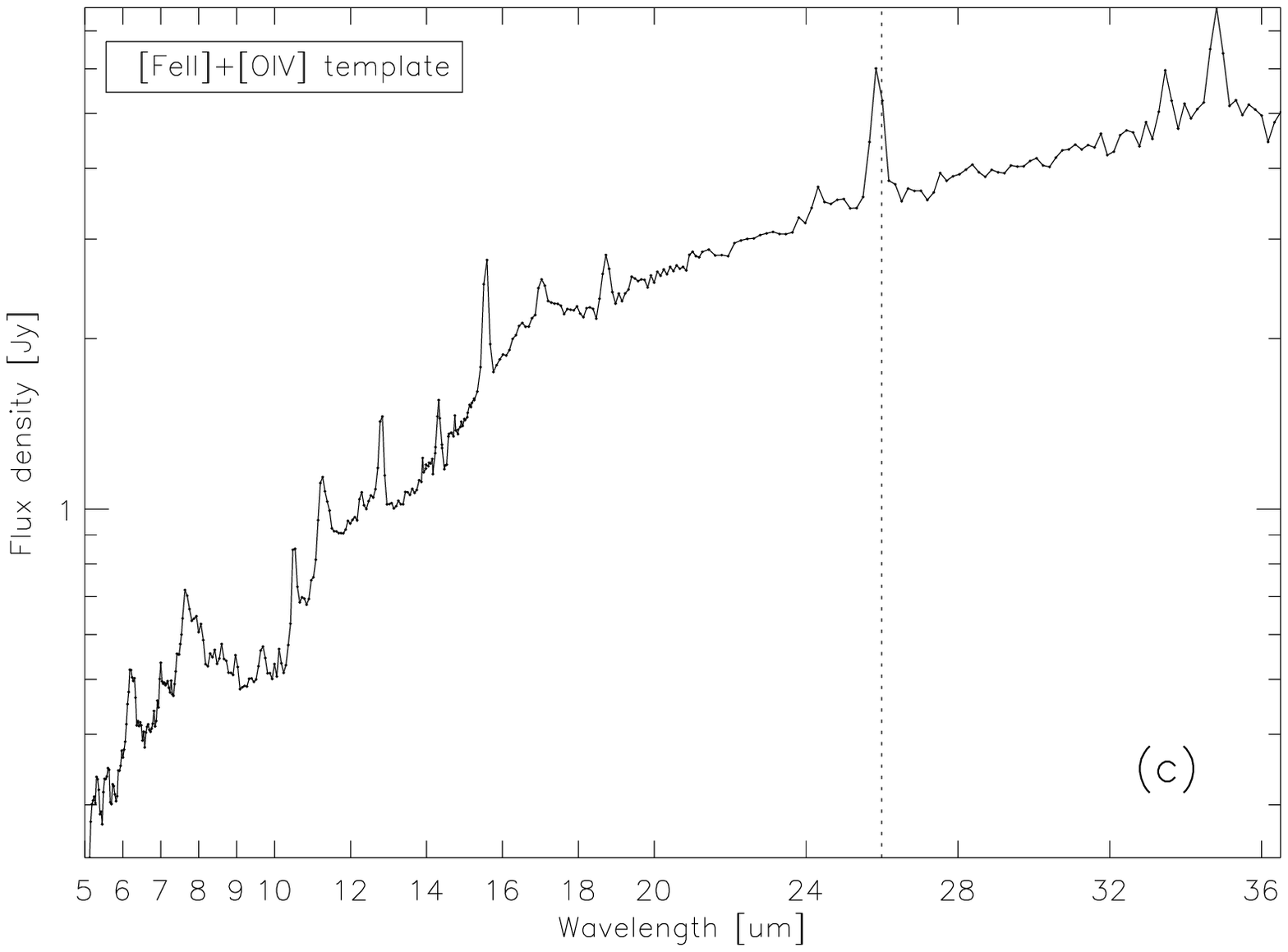}
\includegraphics[angle=0,scale=0.5,clip=true]{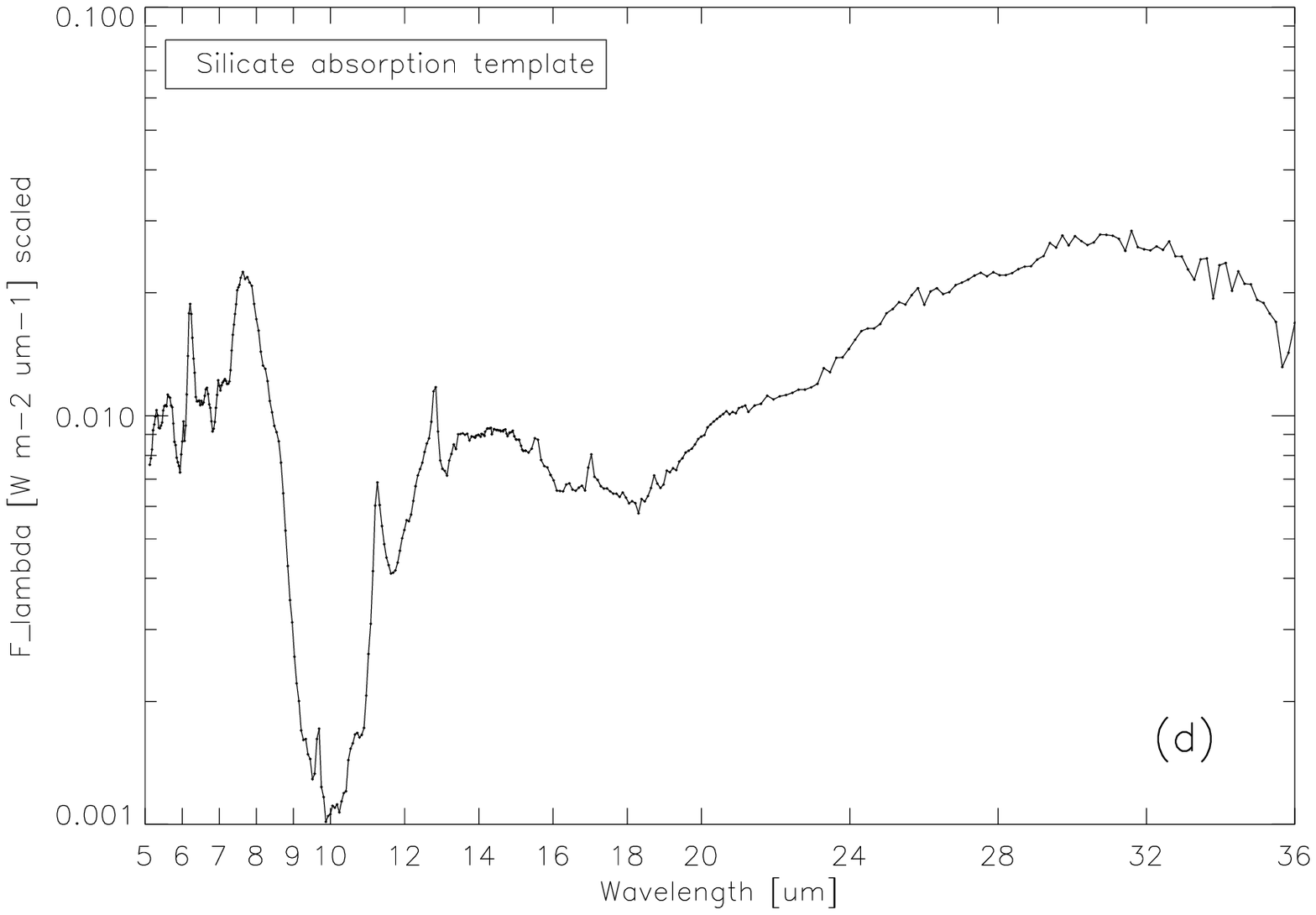}
\caption{Average spectra of extragalactic objects showing significant 6.2\mic\ PAH emission (a), 17.05\mic\ H$_2$ emission (b), [Fe\2]+[O\4] (blended lines) emission (c), and silicate absorption optical depth (d). In each example, the 100 most extreme objects were selected. 
\label{fig:template2}}
\end{figure*}

\begin{figure*}
\includegraphics[angle=0,scale=0.95,clip=true]{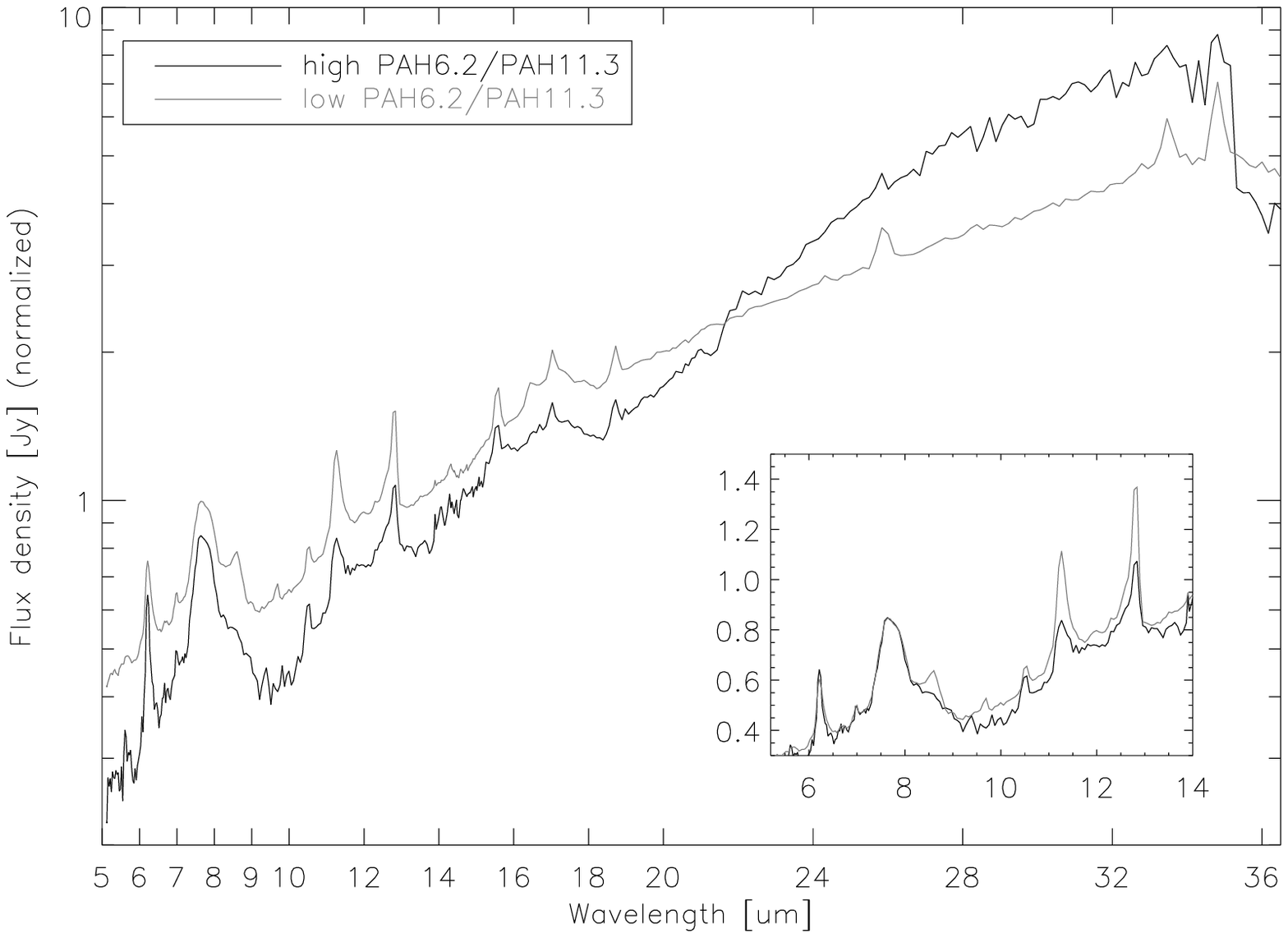}
\caption{Average spectra of extragalactic objects showing the highest and lowest 6.2\mic/11.3\mic\ PAH band ratio. The 100 most extreme cases were selected in each case.
\label{fig:template1}}
\end{figure*}

A few illustrations of extragalactic templates built using spectra in the CASSIS database are shown in Figures\,\ref{fig:template2} and\,\ref{fig:template1}. These spectra can be considered as representative of the objects showing a particular spectral feature. The template  shows the median spectrum  for the 100 sources with strongest features using a given criterion. In these examples, emission from lines and PAHs were simply measured by integrating the flux, no fit was performed. The silicate absorption optical depth was measured according to Spoon et al.\ (2007),  assuming the method usually applicable to PAH-dominated spectra. There is no selection  according to object type, so the spectra include the contribution from starburst galaxies, active galactic nuclei, and all other sources in the CASSIS atlas.

Figure\,\ref{fig:template2} illustrates preliminary investigations for correlating the presence of a given line or band with other spectral  features. The PAH-dominated template agrees well with the starburst template spectra from Brandl et al.\ (2006) and Bernard-Salas et al.\ (2009). Deeper silicate absorption is observed in our template, which is likely due to the contribution from ultraluminous infrared galaxies in addition to starburst galaxies. Another interesting result is the correlation between H$_2$ emission and the [Si\2] emission at 34.82\mic\ (e.g., Roussel et al.\,2007). The objects showing an intense [Fe\2]+[O\4] line complex tend to show strong [Ne\3] emission at 15.56\mic. Finally, the objects with the deepest silicate absorption show few lines and weak PAH bands. 

Figure\,\ref{fig:template1} splits the sample of PAH-dominated sources with the most extreme 6.2\mic/11.3\mic\ PAH ratios. It can be seen that objects with a large 6.2/11.3 ratio show a steeper dust continuum while also having a larger [S\4]/[Ne\2] ratio, suggesting that the PAH band ratio correlates with the radiation field hardness (e.g., Lebouteiller et al.\ 2011; Madden et al.\ 2006; Wu et al.\ 2006; Engelbracht et al.\ 2005). 

\section{Summary}

We present the CASSIS  atlas of $Spitzer$ IRS low resolution spectra. All IRS observations in staring mode are included, totaling about 13\,000 spectra corresponding to 11\,000 distinct sources. The output spectra  provide the best product for a publishable, measurable IRS spectrum. 

An online interface (http://cassis.astro.cornell.edu/atlas) accesses the spectra, provides detailed information on production of the spectra, and allows searches of the atlas based on various parameters.  

Two versions of the spectra are available,  using the optimal extraction provided by AdOpt (suited for point-like sources) and the tapered column extraction (better suited for partially extended sources). Several important diagnostics are  provided, most notably a quantitative detection level and the spatial source extent. 

Future versions of CASSIS will include point-like sources observed serendipitously (at other positions in the slit or within the offset slit),  sources observed with spectral mappings, and the IRS high-resolution observations. 

\acknowledgments 
We wish to thank the people who contributed to the data reduction efforts over the IRS mission. Former ISC members are especially acknowledged (in particular D.\,Devost, D.\,Levitan, D.\,Whelan, K.\,Uchida, J.D.\,Smith, E.\,Furlan, M.\,Devost, Y.\,Wu, L.\,Hao, B.\,Brandl, S.J.U.\ Higdon, P.\ Hall) for their work on the SMART software and for the development of reduction techniques. Moreover, the Rochester group (in particular M.\,McCLure, C.\,Tayrien, I.\,Remming, D.\,Watson, and W.\,Forrest) played an important role in collaborating with the ISC to bring additional and essential improvements to the data reduction used in CASSIS. This research has made use of the NASA/IPAC Extragalactic Database (NED) which is operated by the Jet Propulsion Laboratory, California Institute of Technology, under contract with the National Aeronautics and Space Administration. This research has made use of the SIMBAD database, operated at CDS, Strasbourg, France. This research was conducted with support from
the NASA Astrophysics and Data Analysis Program (Grant NNX10AD61G). Finally, we thank the referee for his/her useful comments.

\appendix
\appendixpage

\section{Acronyms and abbreviations}

\noindent
\textbf{AdOpt}  - ``Advanced Optimal extraction''. Plugin program in the SMART environment notably enabling optimal extraction of Spitzer/IRS spectra by using a super-sampled point spread function.
\\
\noindent
\textbf{AORkey  } (or AOR) - ``Astronomical Observation Request key''. Unique identifier for observations performed by Spitzer. There can be several AORkeys per object.
\\
\noindent 
\textbf{BCD }- ``Basic Calibrated Product''. Calibrated IRS detector images ($128\times128$ pixels) provided by the SSC pipeline. 
\\
\noindent 
\textbf{ BMASK }- ``Bit-mask''. Image plane ($128\times128$ pixels) containing a possible error condition code for each pixel in a detector of the IRS. The bit codes can be found in this page: http://isc.astro.cornell.edu/SmartDoc/ErrorProcessing.
\\
\noindent 
\textbf{ DCE }- ``Data Collection Event''. Single exposure images also referred to as ``ramps''. The ramp duration is the time between the first and last non-destructive reads of the array. Several ramp times are available for each module.
\\
\noindent 
\textbf{ IRS} - ``InfraRed Spectrograph''. The spectrograph onboard the Spitzer Space Telescope.
\\
\noindent
\textbf{ LL} - ``Long-Low''. One of the low spectral resolution modules of the IRS (Table\,\ref{tab:modules}). 
\\
\noindent
\textbf{ PSF }- ``Point Spread Function''. Instrument spatial response function to a point-like source.
\\
\noindent
\textbf{ RSRF} - ``Relative Spectral Response Function''. Function giving the ratio of the observed spectra of a calibration star in uncalibrated units to the theoretical stellar template.
\\
\noindent
\textbf{ SMART }- ``Spectroscopic Modeling Analysis and Reduction Tool''. Contributed software allowing the community to reduce and analyze data from the IRS.
\\
\noindent
\textbf{ SPICE} -  ``SPitzer IRS Custom Extractor''. Official software developed by the SSC to reduce IRS data.
\\
\noindent
\textbf{ SSC} - ``Spitzer Science Center''. Organization at CalTech supporting the science for Spitzer. 
\\
\noindent
\textbf{ SL} - ``Short-Low''. One of the low spectral resolution modules of the IRS (Table\,\ref{tab:modules}).

\end{document}